\documentclass[twocolumn,epjc3,final]{svjour3}          
\RequirePackage[T1]{fontenc}
\RequirePackage{mathptmx}      
\RequirePackage{flushend}
\RequirePackage[colorlinks,citecolor=blue,urlcolor=blue,linkcolor=blue]{hyperref}
\smartqed  

\usepackage{doi}
\usepackage{graphicx}
\usepackage{cite} 
\usepackage{amssymb}
\usepackage{amsmath}
\usepackage{microtype}
\usepackage{siunitx}
\usepackage{color}

\usepackage{upgreek}
\usepackage[utf8]{inputenc} 
\usepackage{comment}

\newcommand{\Eqref}[1]{Eq.~\eqref{#1}}
\newcommand{\Figref}[1]{Fig.~\ref{#1}}

\newcommand{\sub}[1]{_\mathrm{#1}}
\newcommand{\up}[1]{\mathrm{#1}}
\newcommand{\order}[1]{\mathcal{O}(#1)}
\newcommand{\diff}{\mathrm d}

\newcommand{\mean}[1]{\left<#1\right>}

\title{keV-Scale Sterile Neutrino Sensitivity Estimation with Time-Of-Flight Spectroscopy in KATRIN using Self-Consistent Approximate Monte Carlo}
\author{Nicholas M.N. Steinbrink\thanksref{e1, addr1}
        \and
        Jan D. Behrens\thanksref{addr1, addr2} 
        \and
        Susanne Mertens\thanksref{addr3} 
        \and
        Philipp C.-O. Ranitzsch\thanksref{addr1} 
        \and
        Christian Weinheimer\thanksref{addr1}
}
\titlerunning{keV-Scale Sterile Neutrino Sensitivity Estimation with Time-Of-Flight} 
\authorrunning{Nicholas Steinbrink et al.}
\thankstext{e1}{e-mail: nicholas.steinbrink@uni-muenster.de}

\institute{Institut für Kernphysik, WWU Münster, Wilhelm Klemm-Str. 9, 48149 Münster, Germany\label{addr1}
          \and
          Institute of Experimental Particle Physics (ETP), Karlsruhe Institute of Technology (KIT), Wolfgang Gaede-Str. 1, 76131 Karlsruhe, Germany\label{addr2}
          \and
          Physics Department, TU München, James-Franck-Str. 1, 85748 Garching, Germany\label{addr3}
}

 \journalname{Eur. Phys. J. C}
 
\begin{document}
 
\maketitle

\begin{abstract}

We investigate the sensitivity of the Karlsruhe Tritium Neutrino Experiment (KATRIN) to keV-scale sterile neutrinos, which are promising dark matter candidates. Since the active-sterile mixing would lead to a second component in the tritium $\upbeta$-spectrum with a weak relative intensity of order $\sin^2\theta \lesssim \num{e-6}$, additional experimental strategies are required to extract this small signature and to eliminate systematics. A possible strategy is to run the experiment in an alternative time-of-flight (TOF) mode, yielding differential TOF spectra in contrast to the integrating standard mode. In order to estimate the sensitivity from a reduced sample size, a new analysis method, called self-consistent approximate Monte Carlo (SCAMC), has been developed. The simulations show that an ideal TOF mode would be able to achieve a statistical sensitivity of $\sin^2\theta \sim \num{5e-9}$ at one $\sigma$, improving the standard mode by approximately a factor two. This relative benefit grows significantly if additional exemplary systematics are considered. A possible implementation of the TOF mode with existing hardware, called gated filtering, is investigated, which, however, comes at the price of a reduced average signal rate.

\end{abstract}

\section{Introduction}

In recent years the interest has grown for sterile neutrinos with a mass scale of a few keV \cite{WhitePaperWDM2016}. They are proposed as dark matter particle candidates in cold dark matter (CDM) and especially warm dark matter (WDM) scenarios \cite{Destri2013, Destri2013b, Canetti2013, DeVega2012}. WDM has the potential to avoid issues regarding structure formation on small scales which are not yet solved for WIMP (weakly interacting massive particle) CDM \cite{Menci2012, Lovell2012, Evans2009, Schneider2012, Destri2013a, Papastergis2014a, DeVega2014}. However, the shortcomings of WIMP CDM can possibly be mitigated via Baryonic feedback \cite{Chan2015} while any sterile neutrino dark matter production mechanism needs to be fine-tuned to yield the correct DM density. Mass-dependent bounds on the sterile neutrino mixing with active neutrinos have been established by searches for sterile neutrino decay via X-ray satellites \cite{Boyarsky2006, Watson2012} and on basis of theoretical considerations in order to avoid dark matter overproduction \cite{Dodelson1994}, which never exceed $\sin^2 \theta \lesssim \num{e-7}$. The mass range has been constrained by the DM phase-space distribution in dwarf spheroidal galaxies \cite{Boyarsky2009} and gamma-ray line emission from the Galactic center region \cite{Yuksel2008} to $\SI{1}{keV} < m_h < \SI{50}{keV}$. In order to produce the existing amount of dark matter, mass and mixing angle are linked by the production mechanism, which can be non-resonant \cite{Dodelson1994, Shakya2016, Merle2016} or resonant \cite{Shi1999, Shaposhnikov2008, Laine2008, Schneider2016}. Moreover, possible evidence of relic sterile neutrinos with mass $m_h = 7$ keV has been reported in XMM-Newton data \cite{Bulbul2014, Boyarsky2014, Merle2015a}. 

In principle, it can also be searched for keV-scale sterile neutrinos in ground-based experiments,  such as in tritium $\upbeta$-decay {\cite{Shrock1980, DeVega2013}. A promising example is the Karlsruhe Tritium Neutrino Experiment (KATRIN) \cite{KATRINDesignReport}, which is the most sensitive neutrino mass experiment currently under construction. Sterile neutrinos would be visible by a discontinuity in the $\upbeta$-decay spectrum if they have a sufficiently large mixing angle with electron neutrinos. In order to adapt KATRIN, which is optimized for light neutrinos of $m_l \lesssim \order{\si{eV}}$, for keV sterile neutrinos, different approaches are discussed with the goal of enhancing statistics and managing systematics. A suitable idea is to develop a dedicated detector measuring in differential mode \cite{Mertens2015, Mertens2015a, Dolde2017}. As an alternative idea, it is worthwhile to study the performance of an alternative Time-of-Flight (TOF) mode, which has already shown to be promising in theory for active neutrino mass measurements \cite{Steinbrink2013}. 

In this publication the sensitivity of a keV-scale sterile neutrino search based on TOF spectroscopy with the KATRIN experiment is discussed both for an ideal measurement method as for a possible implementation with minimal hardware modifications. 

\section{Sterile Neutrino Search with TOF spectroscopy}

\subsection{Sterile Neutrinos in Tritium \texorpdfstring{$\upbeta$}{Beta}-Decay and KATRIN}

There has been some previous work on sterile neutrinos in general in tritium $\upbeta$-decay. Most publications focus on eV-scale sterile neutrinos \cite{Formaggio2011, Riis2011, Esmaili2012, Kraus2013, Gariazzo2017}, which are proposed to address certain anomalies in oscillation experiments \cite{LSNDCollaboration1998, MiniBooNECollaboration2007, SAGECollaboration2009,  Kaether2010, Mention2011, Giunti2011}. However, in recent time also dedicated studies, dealing with keV-scale neutrinos have been published, such as \cite{DeVega2013, Mertens2015, Mertens2015a}, as well as studies involving more exotic models, such as \cite{Basto-Gonzalez2013, Barry2014c, Steinbrink2017a}. We will quickly summarize the main effect of a keV-scale sterile neutrino on the tritium $\upbeta$-spectrum, while we refer especially to \cite{Mertens2015} for deeper insights into systematics and theoretical corrections.

The tritium $\upbeta$-decay spectrum with a single neutrino with mass eigenstate $m_i$ is given as

\begin{equation}	
\begin{split}
\label{eq:beta1nu}
	\frac{\diff \Gamma}{\diff E} \ = \ & N \frac{ G_\up F^2}{2 \pi^3 \hbar ^7 c^5} \cos^2(\theta_\up C) |M|^2 \ 
	F (E, Z') \cdot p \cdot (E+m_\up e c^2)  \\
	& \cdot \sum\limits_j P_j \cdot (E_0 - V_j - E) \cdot \sqrt{(E_0 - V_j - E)^2 - m_i^2 c^4} ,
\end{split}
\end{equation}

\noindent \cite{Shrock1980, Otten2008, Drexlin2013}, where $E$ is the kinetic electron energy, $\theta_C$ the Cabbibo angle, $N$ the number of tritium atoms, $G_F$ the Fermi constant, $M$ the nuclear matrix element, $F (E, Z')$ the Fermi function with the charge of the daughter ion $Z^{'}$, $p$ the electron momentum, $P_j$ the probability to decay to an excited electronic and rotational-vibrational state with excitation energy $V_j$ \cite{Saenz2000, Doss2006, Doss2008} and $E_0$ the beta endpoint, i.e. the maximum kinetic energy in case of $m_i= 0$. 

The electron neutrino is a superposition of multiple mass eigenstates. Since the flavor eigenstate is the one which defines the interaction, but the mass eigenstate the one which describes the dynamics of the decay, the $\upbeta$-spectrum for the electron neutrino is an incoherent superposition of the contributions for each mass eigenstate,

\begin{eqnarray}
\label{eq:betaspec}
	\frac{\diff \Gamma}{\diff E}(m_{\upnu_\up e}) = \sum_{i=1}^3 |U_{\up ei}|^2 \frac{\diff \Gamma}{\diff E}(m_i) \ .
\end{eqnarray}

In case of an additional keV-scale sterile neutrino, a fourth mass state $m_4$ is introduced with a significantly lower mixing with the electron neutrino, $|U_{\up e4}|^2 \ll |U_{\up ei}|^2 \ (i \in 1, 2, 3)$. In the following we define the \emph{heavy} or {sterile neutrino mass} as $m_h \equiv m_4$  and the \emph{active-sterile mixing angle} as $\sin^2\theta \equiv |U_{\up e4}|^2 < \num{e-7}$ \cite{Watson2012}. Since the light mass eigenstates 1, 2, 3 are not distinguishable by KATRIN \cite{Drexlin2013}, a \emph{light neutrino mass} is defined as $m_l^2 \equiv \sum_{i=1}^3  |U_{\up ei}|^2 m_i^2$. The combined $\upbeta$-spectrum with sterile and active neutrino can then be expressed as

\begin{eqnarray}
	\frac{\diff \Gamma}{\diff E}(m_{\upnu_\up e}) = \sin^2\theta \frac{\diff \Gamma}{\diff E}(m_h) +  \cos^2\theta \frac{\diff \Gamma}{\diff E}(m_l)\ .
	\label{eq:sterilebetaspec}
\end{eqnarray}

\begin{figure}
  \includegraphics[width=\linewidth]{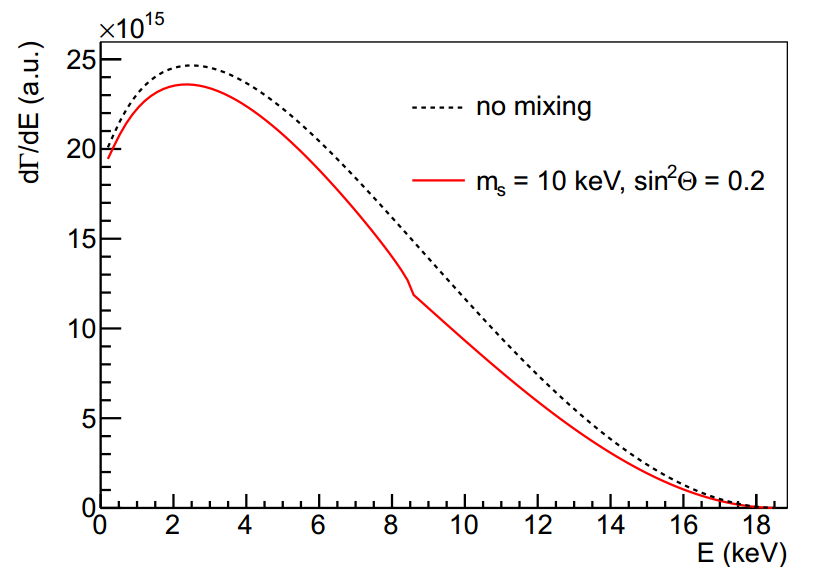}
  \caption{Tritium $\upbeta$-decay spectrum without sterile neutrino contribution (dashed) and with exemplary case of exaggerated mixing with $\sin^2\theta = 0.2 $ and $m_h = \SI{10}{keV}$ (red solid). Figure reproduced from Ref. \cite{Mertens2015}. }
\label{fig:betasterile}
\end{figure}

An example with exaggerated mixing is shown in \Figref{fig:betasterile}. In probing the absolute neutrino mass scale, the KATRIN experiment is designed to measure the light neutrino mass $m_l$ with a sensitivity of $<$ \SI{0.2} {eV}  at 90\% confidence level (CL) \cite{KATRINDesignReport}. Therefor it uses a windowless gaseous molecular tritium source (WGTS) \cite{Babutzka2012} with an activity of $\sim$ \SI{e11}{Bq}. The electrons from the $\upbeta$-decay are filtered in the main spectrometer based on the \emph{magnetic adiabatic collimation with electrostatic filter (MAC-E-Filter)} principle \cite{Picard1992}. The magnetic field in the center of the main spectrometer, the \emph{analyzing plane}, is held small at $B_A = \SI{3}{mT}$ and otherwise high at $B_S = \SI{3.6}{T}$ in the source and at $B_\up{max} = \SI{6}{T}$ at the exit of the main spectrometer just before the counting detector. Due to adiabatic conservation of the relativistic magnetic moment, electron momenta are aligned with the field in the analyzing plane. By additionally applying an electrostatic retarding potential $qU$ in the analyzing plane, the MAC-E-Filter acts as a high-pass filter with a sharp energy resolution of $\Delta E / E = B_A / B_\mathrm{max} \approx \SI{0.9}{eV} / E_0$.  In the focal plane detector (FPD) the count rate is then measured. That way, KATRIN measures the \emph{integral $\beta$-spectrum} as a function of $qU$.

\subsection{Time-Of-Flight Spectroscopy}
\label{sec:tofmode}

The idea of using Time-Of-Flight (TOF) spectroscopy for a measurement of the light neutrino mass is explained in detail in Ref. \cite{Steinbrink2013}. In the following, we will recapitulate the approach briefly and explain the motivations for investigating this technique for a keV-scale sterile neutrino search as well.

In contrast to the standard mode of operation, as described in the last section, TOF spectroscopy allows to measure not only the count-rate, but a full TOF spectrum at a given retarding potential $qU$. The TOF as a function of the energy is given by integrating the reciprocal velocity over the center of motion, which we will assume for simplicity to be on the $z$-axis,

\begin{equation}
	\label{eq:tof}
	\mathcal T (E, \vartheta) = \int \diff z \ {1 \over v_\parallel} 
	= \int\limits_{z_\mathrm{start}}^{z_\mathrm{stop}} \diff z \ \frac{E + m_e c^2 - q \Delta U(z)}{p_\parallel(z) \cdot c^2}\ ,
\end{equation}

\noindent where $E$ and $\vartheta$ are the initial kinetic energy and polar angle of the electron, respectively. $z_\mathrm{start}$ and $z_\mathrm{stop}$ are the positions on the beam axis between which TOF is measured, $\Delta U(z)$ is the potential difference as a function of position $z$ and $p_\parallel(z)$ the parallel momentum. By assuming adiabatic conservation of the magnetic moment, $p_\parallel(z)$ can be expressed analytically as a function of the potential $\Delta U(z)$ and magnetic field $B(z)$ (derivation see Ref. \cite{Steinbrink2013}). If these are known, the integral in \Eqref{eq:tof} can be solved numerically. 

Since the TOF is a function of the energy, the $\upbeta$-spectrum can be transformed into a TOF spectrum $\up d N / \up d \tau$, given the initial angular distribution of the $\upbeta$-decay electrons. A feature in the $\upbeta$-spectrum such as a sterile neutrino contribution would then also have a corresponding effect on the TOF spectrum if the retarding energy $qU$ is sufficiently low. 
Like the $\upbeta$-spectrum \eqref{eq:betaspec}, the TOF spectrum can as well be expressed as a superposition of a component with a heavy neutrino mass $m_h$ and a light neutrino mass $m_l$:

\begin{eqnarray}
\label{eq_betaspec}
	\frac{\diff N}{\diff \tau}(m_{\upnu_\up e}) = \sin^2\theta \frac{\diff N}{\diff \tau}(m_h) +  \cos^2\theta \frac{\diff N}{\diff \tau}(m_l)\ .
	\label{eq:steriletof}
\end{eqnarray}

For each of these two components, the TOF spectrum can then formally be obtained from the $\upbeta$-spectrum with neutrino mass $m_l$ and $m_h$, respectively, using the transformation theorem for densities \cite{Gillespie1983}:

\begin{equation}
  {\diff N \over \diff \tau} = \int_0^{\vartheta_\mathrm{max}} \int_{qU}^{E_0}\diff \vartheta ~ \diff E ~ g(\vartheta) ~ {\diff N \over \diff E}(E, \vartheta) \ \delta\left(\tau - \mathcal T(E, \vartheta)\right) ,
  \label{eq:tofspec}
\end{equation} 

\noindent where $g(\vartheta)$ denotes the angular distribution and ${\diff N}/{\diff E}(E, \vartheta)$ the response corrected energy spectrum, which itself is a function of the $\upbeta$-spectrum \eqref{eq:beta1nu} for a given neutrino mass. If angular changes from inelastic scattering processes in the tritium source are neglected, the angular distribution is approximately independent from the energy spectrum and given by isotropic emission

\begin{equation}
\label{eq:angular}
	g(\vartheta) = {1 \over 2} \sin (\vartheta)
\end{equation}

\noindent within the angular acceptance interval given by the default KATRIN field settings with  $\vartheta_\mathrm{max} = \sqrt{B_S / B_\mathrm{max}} = \SI{50.77}{\degree}$. The response corrected energy spectrum ${\diff N}/{\diff E}(E, \vartheta)$ in \Eqref{eq:tofspec} is given in good approximation by the $\upbeta$-spectrum \eqref{eq:sterilebetaspec}, convolved with the inelastic energy loss function in the tritium source,

\begin{equation}
\label{eq:betaeloss}
\begin{split}
	\frac{\diff N}{\diff E}(E|\vartheta)
	& = 
	{\diff \Gamma \over \diff E} \otimes f\sub{loss}(E, \vartheta) \\
	& =
	p_0(\vartheta) \cdot  {\diff \Gamma \over \diff E} 
	+ 
	\sum\limits_{n=1}^\infty p_n(\vartheta) \cdot {\diff \Gamma \over \diff E} \otimes f_n (E) \
\end{split}
\end{equation}

\noindent where the $f_n$ is the energy loss spectrum of scattering order $n$ which can be approximately defined via recursive convolution through the single scattering energy loss spectrum $f_1$. This can be written as 

\begin{equation}
\label{eq:eloss}
f_n =  f_{n-1} \otimes f_1 \qquad (n > 1) \ .
\end{equation}

\noindent The probability $p_n$ that an electron is scattered $n$ times depends on the emission angle $\vartheta$ and is given by a Poisson law

\begin{equation}
\label{eq:pscat}
	p_n(\theta) = \frac{\lambda^n(\vartheta)}{n!}e^{-\lambda(\vartheta)}~.
\end{equation}

\noindent The average number of scattering processes $\lambda$ is given in terms of the column density $\rho d$ of the tritium source, the mean free column density $\rho d\sub{free}$ and the scattering cross section $\sigma\sub{scat}$ as
\begin{equation}
	\lambda(\vartheta) = \int_0^1 \diff x \ \frac{\rho d \cdot x}{\rho d \sub{free} \cdot \cos \vartheta} = \int_0^1 \diff x \ \frac{\rho d \cdot x \cdot \sigma\sub{scat}}{\cos \vartheta}~.
\end{equation}

Since the probability of $n$-fold scattering is a function of the emission angle \eqref{eq:pscat}, the response corrected energy spectrum \eqref{eq:betaeloss} itself becomes dependent on the angle. Note that the scattering model is simplified, since angular changes in collisions are neglected and the scattering probabilities are averaged over a hypothetical uniform density profile in the source. We would like to clarify that in our actual implementation the $n$-fold energy loss spectra are not generated via convolution but via Monte Carlo, which yields, however, equivalent results. Furthermore, using \Eqref{eq:tof}, the radial starting position is always assumed to be $r = 0$, which is not the case in KATRIN, but we do not expect significant changes in the spectral shape for outer radii. For analysis of real experimental data a fully realistic treatment would be necessary, yet for a principle sensitivity study these approximations are reasonable.

\begin{figure}
  \includegraphics[width=1.05\linewidth ]{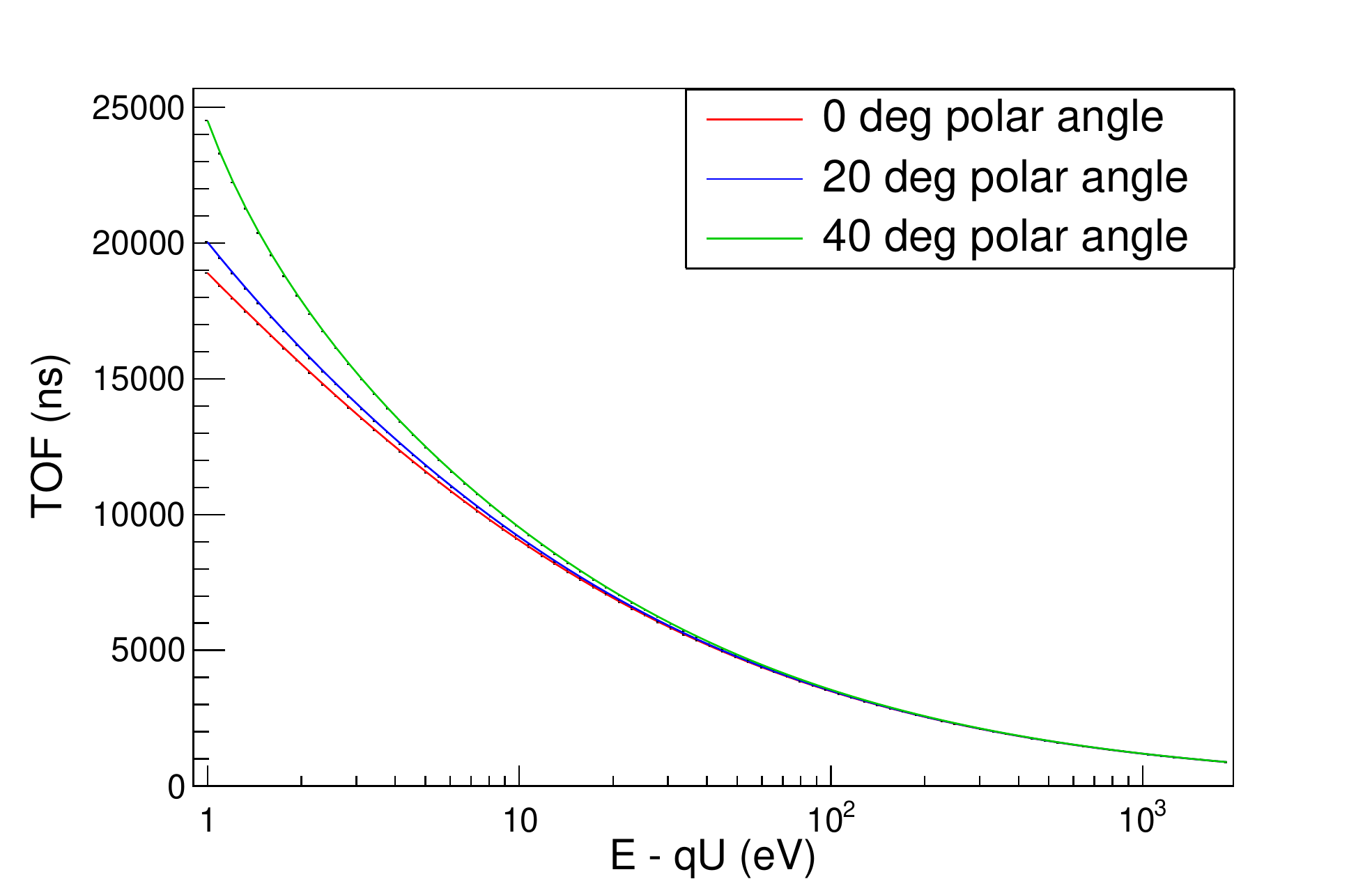}
  \caption{TOF as a function of surplus energy $E-qU$. Significant energy differences are detectable up to a few $\SI{100}{eV}$ above the filter threshold. By combining multiple TOF spectra with different retarding energies, the TOF method will give a differential map of the energy spectrum within the measuring interval. }
  \label{fig:tof-e}
 \end{figure}

The benefits of a TOF measurement can be understood from \Figref{fig:tof-e}, where the TOF \eqref{eq:tof} as a function of $E$ for different angles is shown. It can be seen that energy differences up to some $\sim \SI{100}{eV}$ above the retarding potential translate into significant TOF differences. Within these regions, TOF spectroscopy is thus a sensitive \emph{differential} measurement of the energy spectrum. Combining multiple TOF spectra measured at different retarding energies thus allows to measure a differential equivalent of the $\upbeta$-spectrum throughout the whole region of interest. As already outlined in Ref. \cite{Mertens2015}, a differential measurement has important benefits for a sterile neutrino search. On the one hand it enhances the statistical sensitivity since the sterile neutrino signature can be measured directly without any intrinsic background from higher energies as in the classic high pass mode. On the other hand, it reduces the systematic uncertainty since it improves the distinction between systematic effects and a real sterile neutrino signature in the spectrum.

\subsection{TOF Measurement}

As the approach is rather novel, most existing ideas for TOF measurement are still in an early development phase and have not been tested. There are ongoing efforts to develop hardware which is intended to detect passing electrons with minimal interference with their energy (\emph{electron tagger}) \cite{Steinbrink2013}. Approaches are amongst others to measure tiny excitations induced in an RF cavity or to detect the weak synchrotron emission of the electrons in the magnetic field via long antennas (cf. Refs. \cite{Monreal2009, Project82015}). While promising, there has unfortunately not been any break-through in the technical realization for such an electron tagger, yet. Additionally, it seems unlikely that such an approach is also useful for keV sterile neutrino searches. For a sufficient sensitivity on $\sin^2\theta$ the count-rate needs to be as high as possible. However, count-rates much above $\SI{10}{kcps}$ would lead to ambiguities in the combination of a start signal in the electron tagger and the stop signal in the detector given the overall TOF of order $\sim\si{\micro s}$ (see \Figref{fig:tof-e}). 

\begin{figure}
\centering
  \includegraphics[width=\linewidth]{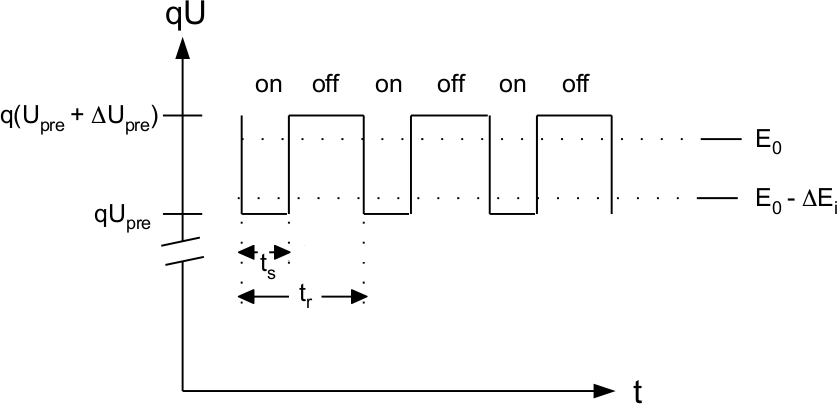}
  \caption{Pre-spectrometer potential pulsing scheme for gated filter. X-axis: time. Y-axis: pre- spectrometer retarding potential. At the lower filter setting all electrons of the interesting region of width $\Delta E_i$ below the endpoint $E_0$ are transmitted while at the higher setting all electrons are blocked.}
  \label{fig:gatedfilter}
\end{figure}

A method which has already been tested in the predating Mainz experiment \cite{Bonn1999} is a periodic blocking of the electron flux, called \emph{gated filtering} (GF). If electrons are only transmitted during a short fraction of the time, the arrival time spectrum would approximate the TOF spectrum. In KATRIN, this could for instance be achieved by pulsing the pre-\-spec\-tro\-meter potential between one setting with full transmission and one setting with zero transmission (\Figref{fig:gatedfilter}). The main downside of the method is that it sacrifices statistics in order to get time information. However, it would require minimal hardware modifications since only the capability to pulse the pre-spectrometer potential by some keV would have to be added. Since the focal plane detector of KATRIN is optimized for low rates near the endpoint, the method could also in principle be utilized for an early keV sterile neutrino search by using a small duty cycle with sharp pulses and thereby reducing the count-rate. However, in this scenario with small hardware modifications, it is unlikely that the pre-spectrometer potential can be pulsed by more than some keV. Due to the capacity of the pre-spectrometer, there is possibly a non-vanishing ramping time involved, depending on the ramping interval. If electrons arrive within the ramping time, they become either accelerated or retarded, giving rise to non-isochronous background. The problem can be mitigated partly by using a voltage supply with higher power. Alternatively, a mechanical high-frequency beam shutter could be use. However, this would come at the cost of larger modifications of the set-up and a lower flexibility regarding fine-tuning of the timing parameters. We will not discuss this problem further and just assume an ideally efficient method of periodically blocking the beam. However, we will restrict the sensitivity study of the sterile neutrino search with the GF method to a measurement region spanning only a few keV below the endpoint.

\section{Monte Carlo Sensitivity Estimation}
\label{sec:sensitivitymethod}

The TOF spectrum \eqref{eq:tofspec} can not be calculated analytically, since the magnetic field $B(z)$ and electron potential $q\Delta U(z)$ are only known numerically. There are two remaining possibilities of simulating TOF spectra. The first approach is to to evaluate the $\delta$ function in the TOF spectrum \eqref{eq:tofspec} via numerical integration. This method has been used in Ref. \cite{Steinbrink2013} since it delivers generally precise results and is well scalable. The bottleneck of this method is, however, the convolution of the $\upbeta$-spectrum with the $n$-fold energy loss spectra \eqref{eq:betaeloss}. The convolution routine is rather performance-intensive especially for a large spectral surplus $E_0-qU$ (as present in case of keV scale sterile neutrino search) and requires complicated optimizations to work successfully. Furthermore, if the addition of further effects such as angular-changing collisions might be requested for future studies, the implementation will become more difficult.

Therefore, we chose to apply the second approach which is to generate the TOF spectra \eqref{eq:tofspec} via Monte Carlo (MC) simulation. This especially avoids the convolution of the $\upbeta$-spectrum with the energy loss function \eqref{eq:betaeloss}, since the energy loss can be randomly generated individually without additional expensive convolutions. While a MC approach is generally very flexible when it comes to the addition of more detailed effects and systematics, it is generally not as scalable in terms of the expected number of events as a purely numerical approach. KATRIN is designed for measurements near the $\upbeta$-endpoint with low rates on the order of several \si{cps}. The measurements for the keV-scale sterile neutrino detection, however, have to be performed over a significantly broader region of the $\upbeta$-spectrum and thus count rates up to $\sim \SI{e10}{cps}$ can be expected. For a data taking period of three years, one would thus expect up to $\sim \SI{e18}{cps}$. If a realistic model for a sensitivity analysis shall be simulated event-by-event, it is obvious that the sample size needs to be significantly larger than the expected number of events. In our case, the calculation of flight times of more than \num{e18} events is simply not possible within a reasonable computing time. 

However, we will show that, if the signal is sufficiently small compared to the total expected rate, the dominating "background part" of the model (corresponding to the $\cos^2\theta$-term in \Eqref{eq:steriletof}) can be approximated. This works due to the fact that for a pure sensitivity study, as opposed to an analysis of real data, only the fidelity of the signal is relevant.

\subsection{Self-Consistent Approximate Monte Carlo}


In this section, we argue that a modified Monte Carlo strategy, from here on called \emph{self-consistent approximate Mon\-te Car\-lo} (SCAMC), will be able to reduce the necessary total sample size in a sufficient amount to address the problems mentioned above. This works if two requirements are met. These are 

\begin{enumerate}
  \item that the model can be separated into a background part and a signal part, with the latter sufficiently smaller than the first, and
  \item that model and toy data are self-consistent, i.e. the toy data are sampled directly from the model.
\end{enumerate}

We will first discuss this approach for a generic case. Assume, the model distribution $\Phi$ can be expressed by a linear combination
 
\begin{equation}
	\Phi = c_\up S \Phi_\up S + c_\up B \Phi_\up B  ,
	\label{eq:scis}
\end{equation}

\noindent consisting of a \emph{signal} contribution $c_\up S \Phi_\up S$, sampled with maximum precision, and an approximated \emph{background} contribution, $c_\up B \Phi_\up B$. The distribution of interest is then replaced by a modified distribution

\begin{equation}
	\Phi' = c_\up S \Phi_\up S + c_\up B \Phi_\up B' , 
	\label{eq:approxmodel}
\end{equation}

\noindent with $ \Phi_\up B' \sim \Phi_\up B$, where the background component is either approximated by an analytic expression or simulated by MC with a reduced sample size. 
We demand that $\Phi_\up B$ is independent of any parameter of interest, $\mu$ (and of any parameter which is strongly correlated with a parameter of interest):

\begin{equation}
	\frac{\mathrm d\Phi_\up B}{\mathrm d\mu} = 0 \ . 
	\label{eq:sigbgcondition}
\end{equation}


The approximate model \eqref{eq:approxmodel} can then be used as replacement for the real model. The sensitivity estimation can then be continued in the standard frequentist way: toy data are sampled from $\Phi'$ for given parameter choices and the confidence region for the parameter of interest $\mu$ can then be determined via $\chi^2$ fits. 

The benefit of this strategy can be understood in the following way. Since the data have been sampled from the model, any error in the model will also be passed over to the data. However, while the total approximated distribution $\Phi'$ itself is inaccurate, it still contains all essential information about the sensitivity, since $\Phi' - c_\up B \Phi'_B = c_\up S \Phi_\up S$ holds exactly (\Figref{fig:scis}). Since only the fidelity of the signal is relevant for the sensitivity analysis (which we assured with condition \eqref{eq:sigbgcondition}), both the error in the model and in the data approximately cancel each other in the fit. It can be shown in this case that the width of the $\chi^2$ minimum stays the same as long as the background component is at least approximately correct. 
A simplified proof can be found in \ref{app:proof}.

\begin{figure}
  \includegraphics[width=1.05\linewidth]{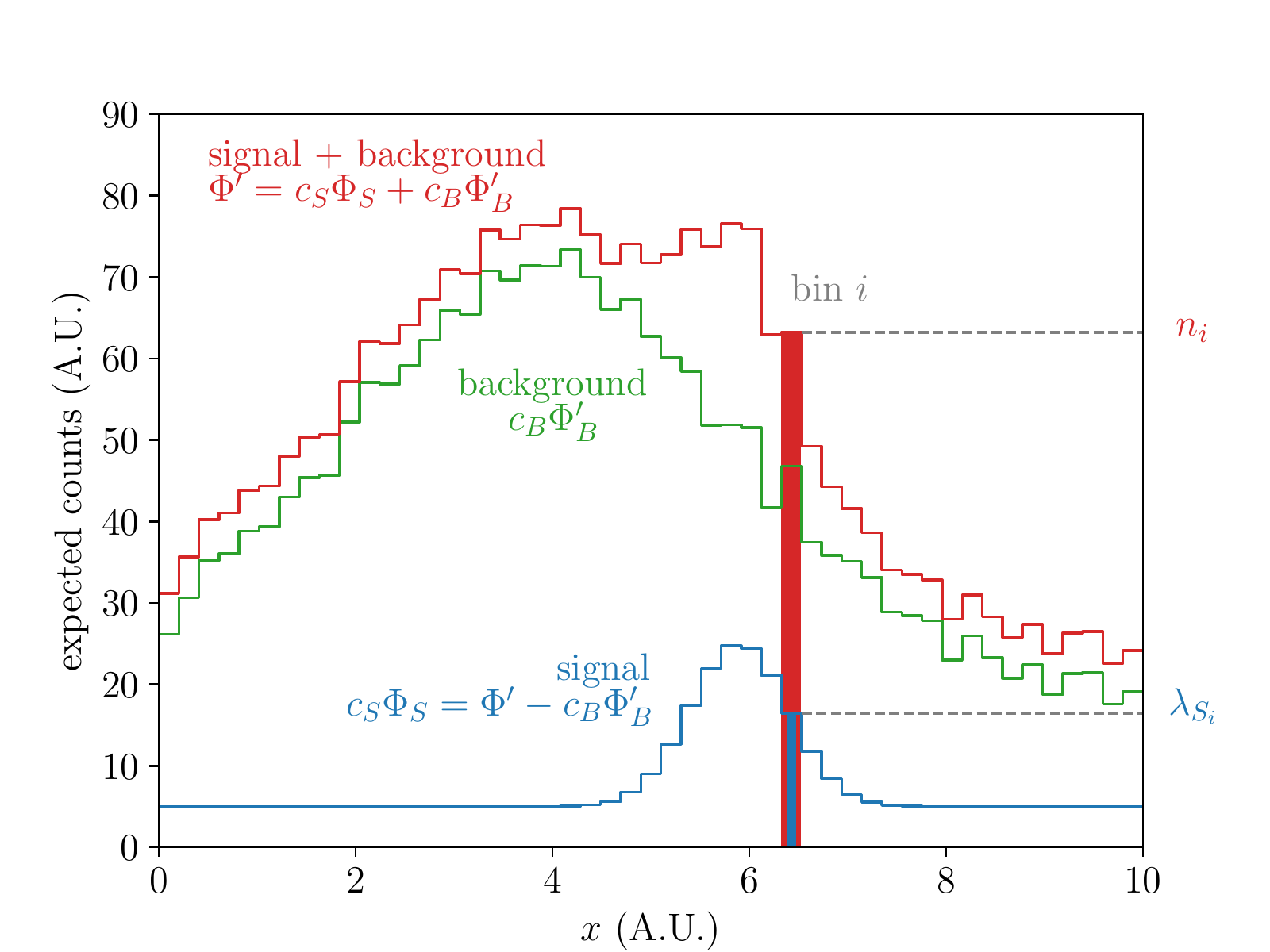}
  \caption{Illustration on the SCAMC approximated model for the example of a sum of two Gaussians. The signal term $c_\up S\Phi_\up S$ is an analytic Gaussian, while the background term $c_\up B\Phi_\up B'$ has been sampled with low statistics MC. The total approximated distribution $\Phi'$ is then inaccurate but contains all essential information about the signal, because $\Phi' - c_\up B \Phi'_B = c_\up S \Phi_\up S$ holds exactly. }
  \label{fig:scis}
\end{figure}

We shall discuss the method now on the initial case of the keV scale sterile neutrino search with TOF spectroscopy. As derived above, the electron TOF spectrum \eqref{eq:tofspec} with added sterile neutrinos can be expressed as a superposition of two TOF spectra with a light or heavy neutrino mass, $m_l$ and $m_h$, respectively. We identify the signal with the sterile neutrino component of the TOF spectrum \eqref{eq:steriletof} and the background with the active neutrino contribution,

\begin{equation}
	\Phi_\up S = \frac{\mathrm d N}{\mathrm d \tau} (m_h) \qquad \Phi_\up B = \frac{\mathrm d N}{ \mathrm d \tau} (m_l) \ .
\end{equation}

\noindent The coefficients are then given by the active-sterile mixing, 

\begin{equation}
	c_\up S  = \sin ^2 \Theta \qquad c_\up B = \cos ^2 \Theta \ .
\end{equation}

It is obvious that for a small signal fraction of, e.g., $\sin^2\theta \lesssim \num{e-6}$, only a small fraction of the total expected events needs to be simulated now. However, since signal and background are always measured together and not separately, the required sample size is reduced even more. For demonstration purposes, let us define the signal expectation value in bin $i$ as 

\begin{equation}
    \lambda_{\up S_i} = n \cdot c_\up S \cdot \Phi_\up S(x = X_i) \ ,
\end{equation}

\noindent with $n$ as total number of expected events (see \Figref{fig:scis}). We will denote the number of expected events in bin $i$ as $n_i$. To approximate the necessary sample size, we require that the numerical uncertainty of $\lambda_{\up S_i}$ needs to be smaller than the expected measurement uncertainty of the number events in the corresponding bin, $\sigma_i$:

\begin{equation}
	 \Delta\lambda_{\up S_i} \ll \ \sigma_i \  ,
	 	\label{eq:errcondition2}
\end{equation}

\noindent Assuming a Poissonian measurement uncertainty, $\sigma_i = \sqrt{n_i}$ and using $\Delta\lambda_{\up S_i}/\lambda_{\up S_i} = 1 / \sqrt{N_\up{S_i}}$, where $N_\up{S_i}$ denotes the signal sample size in bin $i$, \Eqref{eq:errcondition2} gives

 \begin{eqnarray}
	 \frac{1}{\sqrt{N_\up{S_i}}}  \ll \frac{\sqrt{n_i}}{\lambda_{\up S_i}}
	 \iff N_\up{S_i} \gg \frac{\lambda_{\up S_i}^2}{n_i} ,
\end{eqnarray}

\noindent We now define the total signal sample size as  $N_\up S = \sum_i N_\up{S_i}$. If we assume the signal-background ratio to be roughly within a constant order of magnitude, we get the required minimum signal sample size:

\begin{eqnarray}
	 N_{\up S} \gg  \sum_i \frac{\lambda_{\up S_i}^2}{n_i} \approx \frac{n_i^2 \cdot c_\up S^2}{n_i}  = n \cdot c_\up S^2\ .
	 \label{eq:sizecondition2}
\end{eqnarray}
 
\noindent Naively, one would suppose that the signal part still needs to be sampled with full statistics, i.e. $N_{\up S}\gg n \cdot c_\up S$. However, due to the fact that the signal part is always measured with background, we have shown that an additional suppression factor of $c_\up S$ applies. Assuming $\sin^2\theta \sim \num{e-6}$ and a total event size of $n \sim \num{e18}$, we thus get

\begin{equation}
  n \cdot c_\up S^2 = n \cdot \sin^4 \theta \ \sim \ \num{e18} \cdot \num{e-12} = \num{e6} \ .
\end{equation}

\noindent Note that $\sin^2\theta \sim \num{e-6}$ represents roughly the upper bound from astrophysical observations. Likewise, $n \sim \num{e18}$ is approximately the maximum number of counts which will decrease with higher retarding potentials. Thus, for lower values of either one, the necessary sample size is reduced even more according to condition \eqref{eq:sizecondition2}. 



\section{Simulation}

\subsection{Probabilistic model: TOF Spectra}

Using a Monte Carlo algorithm, the TOF spectra given by the transformation \eqref{eq:tofspec} can be determined in a straightforward way. For each MC sample, at first an initial energy and starting angle is generated. The angular distribution is given by \Eqref{eq:angular}. For the initial energy, the electronic excited state is generated from the final state distribution in \Eqref{eq:beta1nu} and then the energy is generated from the respective $\upbeta$-spectrum component in \Eqref{eq:betaspec}. Given the initial energy and the starting angle, the number of inelastic scattering process in the source is generated from \Eqref{eq:pscat} and for each process the energy loss is generated from \Eqref{eq:eloss} and subtracted from the energy. In order to further optimize the Monte Carlo method for a parametrizable heavy neutrino mass, the TOF spectra have additionally been decomposed into elements corresponding to different sterile neutrino mass phase space segments, which is explained in detail in \ref{app:decomposition}. The advantage of such a scheme is that already simulated Monte Carlo events can be reused for different sterile neutrino masses.

We found that a sample size of $10^8$ for each sterile sub-component is feasible in finite calculation time and sufficient for an accurate simulation. The active neutrino component, which contains $\sim 1/\sin^2\theta$ more counts than the total sterile component, was approximated with a sample size of $10^9$, according to the SCAMC approach. The active neutrino mass was set to $m_l = 0$ and the endpoint held constant at $E_0 = \SI{18.575}{keV}$, since there is no correlation to expect with the sterile neutrino. The bin width was chosen to be \SI{250}{ns} (compared to the FPD time resolution of about \SI{50}{ns}) for reasons of performance and robustness. However, it is unlikely to expect for any measurement method to achieve a higher resolution. To all spectra a Gaussian time uncertainty of $\Delta \tau=\SI{50}{ns}$ was added to account for the detector time resolution and a isochronous background of $b=\SI{10}{mcps}$.

\begin{figure}
  \includegraphics[width=\linewidth]{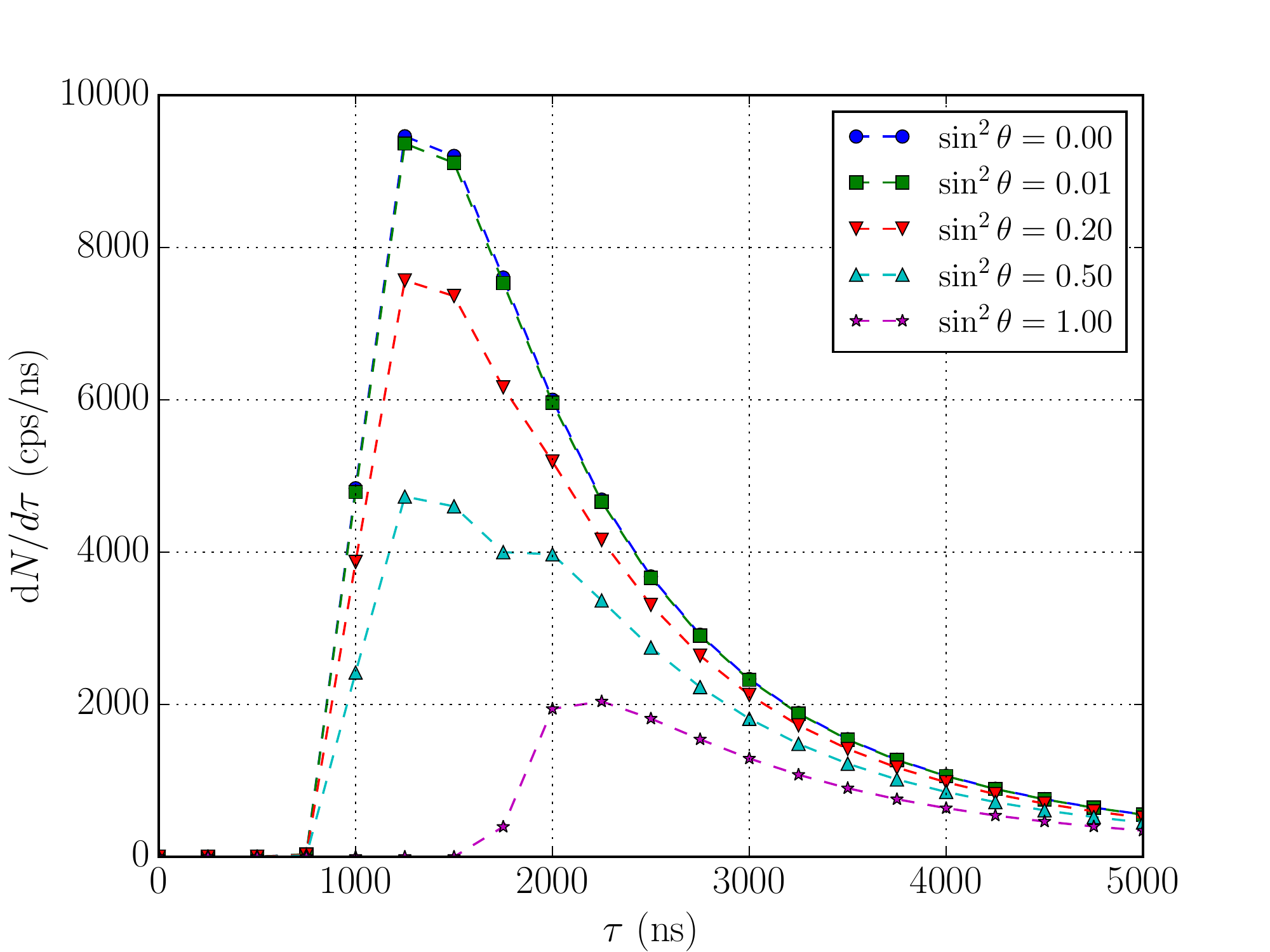}
  \caption{Electron TOF spectra for a keV-scale sterile neutrino of $m_h = \SI{1.1} {keV}$ and different mixing angles at a fixed retarding potential of \SI{17}{keV}. The mixing angles have been exaggerated to enhance the signature and comprise additionally the case of no mixing ($\sin^2 \theta = 0$) as well of pure sterile contribution ($\sin^2 \theta = 1$). Similar to the tritium $\upbeta$-decay energy spectrum, the signature of a sterile neutrino is a kink-like discontinuity at a certain point in the TOF spectrum. Figure first published in \cite{WhitePaperWDM2016}.}
  \label{fig:tofspec-mixings}
  \end{figure}

\begin{figure}
  \includegraphics[width=\linewidth]{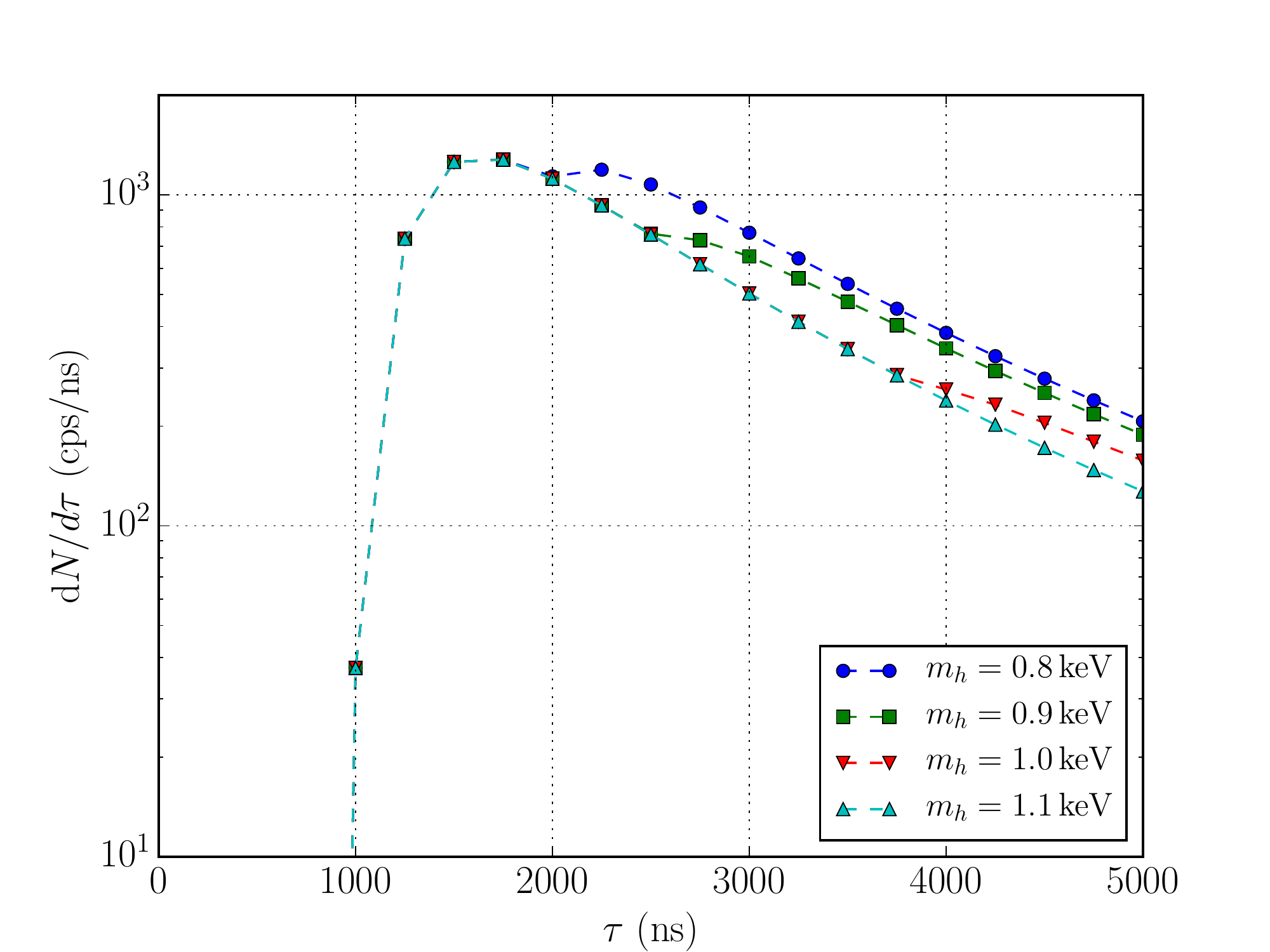}
  \caption{Electron TOF spectra for different sterile neutrino masses at a fixed retarding potential of \SI{17}{keV}. The mixing has been set to $\sin^2 \theta = 0.5$ to enhance the signature. The heavy neutrino mass determines the position of the kink on the TOF-axis. The on-set TOF for a certain sterile neutrino mass can be estimated from \Figref{fig:tof-e}. }
  \label{fig:tofspec-masses}
\end{figure}

Figs. \ref{fig:tofspec-mixings} and \ref{fig:tofspec-masses} show exemplary simulated TOF spectra for different active-sterile mixings and heavy neutrino masses, respectively.  It can be seen that the spectra show a dominating peak within the first \SI{2}{\micro s} which consists of the fast electrons more than some 100 eV above the retarding potential. They are, however, followed by a long tail where the electron velocity becomes slower and the TOF difference per given energy difference (see \Figref{fig:tof-e}) becomes more significant. In this region the TOF spectrum is to a good extent a differential representation of the $\beta$ spectrum, while the fast peak region consists only of some bins, thus contributing to the sensitivity more by its integral. If the sterile neutrino mass is some 100 eV smaller than the difference of endpoint and retarding potential, the sterile neutrino signal becomes similar to that one in the tritium beta spectrum. The sterile neutrino contribution appears as a discontinuity in shape of a "kink" at a certain position in the spectrum. Since the relationship between energy and TOF is non-linear, the position of the kink allows no direct analytical conclusion about the sterile neutrino mass. However, given the retarding potential, the relation in \Figref{fig:tof-e} can be used for an estimation.

\subsection{Ideal TOF mode Sensitivity}

The model described in the last section was utilized to estimate the sensitivity according to the procedure described in section \ref{sec:sensitivitymethod}. The fits   have generally been performed by a $\chi^2$ minimizations using MINUIT \cite{James1975} . For statistical sensitivity estimation, the mixing $\sin^2\theta$ and overall amplitude $S$ are free fit parameters, using a range of fixed values for $m_h$. In those simulations, where the uncertainty on $m_h$ is of interest, also the squared heavy neutrino mass $m_h^2$ has been included as fit parameter. Since each fit incorporates a set of multiple measurements at different retarding potentials,  the $\chi^2$ functions of each measurement are added and fitted with global fit parameters. Instead of a pure ensemble approach, the parameter uncertainties have been calculated using the module MINOS from MINUIT \cite{James1975}, averaged over multiple simulations, which gives in case of an approximately quadratic $\chi^2$ near the minimum an identical result.

\subsubsection*{Exemplary Systematics}

\label{sec:systematics}

In addition to the statistical sensitivity, an exemplary systematic effect has been studied, which is the inelastic scattering cross section due to fluctuation in the column density as described in \Eqref{eq:eloss}. This is one of two main systematics when it comes to keV sterile neutrino search, the other being the final state distribution \cite{Saenz2000, Doss2006, Doss2008}. To incorporate the systematics, the $\chi^2$ function has been modified by an additional term:

\begin{equation}
  \chi^2 = \chi_0^2 + \frac{(\rho d - \mean{\rho d} ) ^2}{(\Delta \rho d)^2} \ ,
  \label{eq:pullterm}
\end{equation}

\noindent where $ \chi_0^2$ is the default binned $\chi^2$ function, $\rho d$ the fitted column density, $\mean{\rho d}$ its expectation value and $\Delta \rho d$ the systematic uncertainty. In order to be able to have $\rho d $ as free fit parameter, the complete model has additionally been separated by number of inelastic scattering processes and weighted with the $l$-fold energy loss probability $p_l(\rho d)$ as given by \Eqref{eq:pscat}, instead of randomly generating the number of inelastic scattering events,

\begin{equation}
  \frac{\mathrm d N}{\mathrm d \tau} = \sum_l p_l(\rho d) \cdot \left( \frac{\up d N}{\up d \tau}\right)_l \ . 
\end{equation}

To determine the influence of the uncertainties $\Delta \rho d$ on the sensitivity, the column density has been shifted by for the data generation by $\rho d = \mean{\rho d} + \Delta \rho d$ while still using the unshifted expectation value $\mean{\rho d}$ in \Eqref{eq:pullterm}. By this approach the MINOS error will increase plus a possibly slight bias in average which is then quadratically added to the average error bars. 

To illustrate the imprint of the systematic uncertainty of $\rho d$ in the TOF spectrum, \Figref{fig:systematics} shows the difference between a TOF spectrum with shifted column density, $\Phi(\rho d) = \diff N / \diff \tau (\rho d + \Delta \rho d)$ and a TOF spectrum with mean column density,  $\Phi_0 = \diff N / \diff \tau (\mean{\rho d})$, weighted by $\sqrt{\Phi_0}$ which is proportional to the expected Poissonian uncertainty of the data. By doing so, the signature becomes visible proportionally to its impact in the $\chi^2$ function. It can be seen that the imprint of a shifted column density is present foremost at lower flight times, which is since the energy loss causes the count-rate near the endpoint to drop. There are fluctuations at higher flight times near the retarding potential arising from the energy loss spectrum \eqref{eq:eloss}. However, these are weighted minimally since the differential rate in the TOF spectrum drops with higher flight times (see \Figref{fig:tofspec-mixings}). 

\begin{figure}
  \includegraphics[width=1.05\linewidth]{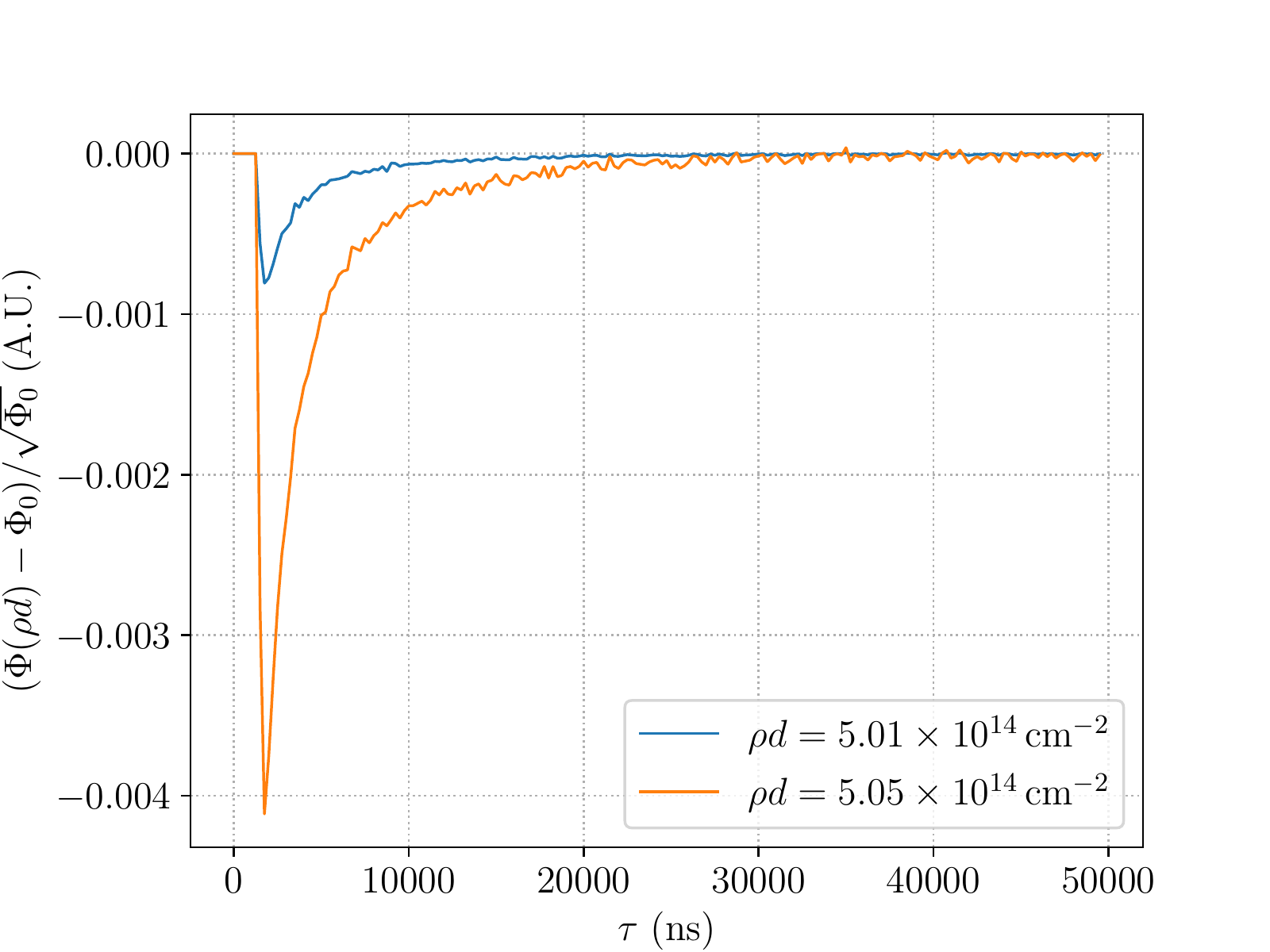}
  \caption{Difference between TOF spectra with shifted $\rho d$, $\Phi(\rho d)$ and default value $\mean{\rho d} = \SI{5e14}{cm^{-2}}$, $\Phi_0$, weighted proportionally with the expected Poissonian uncertainty of the data $\propto \sqrt{\Phi_0} $. The imprint of a shifted column density is present foremost at lower flight times, due to missing events near the endpoint because of the energy loss. Fluctuations at higher flight times near the retarding potential are suppressed by a lower differential count rate. The spectra consist only of the active neutrino component, $\sin^2\theta = 0$, and the retarding potential is $qU = \SI{18}{kV}$.}
  \label{fig:systematics}
\end{figure}

\subsubsection*{Results}

\Figref{fig:idealtofsensitivity} shows the sensitivity for an ideal TOF mode. The results are based on three years measurement time which was distributed uniformly on the retarding potential within an interval of [4; 18.5] \si{keV} with steps of \SI{0.5}{keV}. The setting was chosen in that way that a 7 keV neutrino signal \cite{Bulbul2014} would roughly lie in the center of the potential distribution. For the exemplary inelastic scattering systematics an initial uncertainty of $\Delta\rho d / \rho d = 0.002$ has been assumed in accordance with Ref. \cite{KATRINDesignReport}. The statistical sensitivity of the integral mode in this simulation is in good agreement with Ref. \cite{Mertens2015}. The statistical sensitivity of the ideal TOF mode is close to that one of an ideal differential detector in the aforementioned publication. However, if the uncertainties of the column density are incorporated, the benefit by the TOF mode grows even further, since a shifted column density has a unique imprint on the TOF spectrum (see \Figref{fig:systematics}), which is not the case in the integral mode.  

It should be noted, however, that for low retarding potentials as used in \Figref{fig:idealtofsensitivity}, adiabaticity of the electron transport is limited. Yet, that can be maintained by increasing the magnetic field in the main spectrometer. This lowers the energy resolution and thus the transformation of transverse into longitudinal momentum, which would manifest in a stronger angular-dependence of the energy-TOF relation in \Figref{fig:tof-e}. Though, this should have no significant influence on the sensitivity since the measurement takes place on a keV-scale where the requirements for magnetic adiabatic collimation are more relaxed. 

\begin{figure}
  \includegraphics[width=1.05\linewidth]{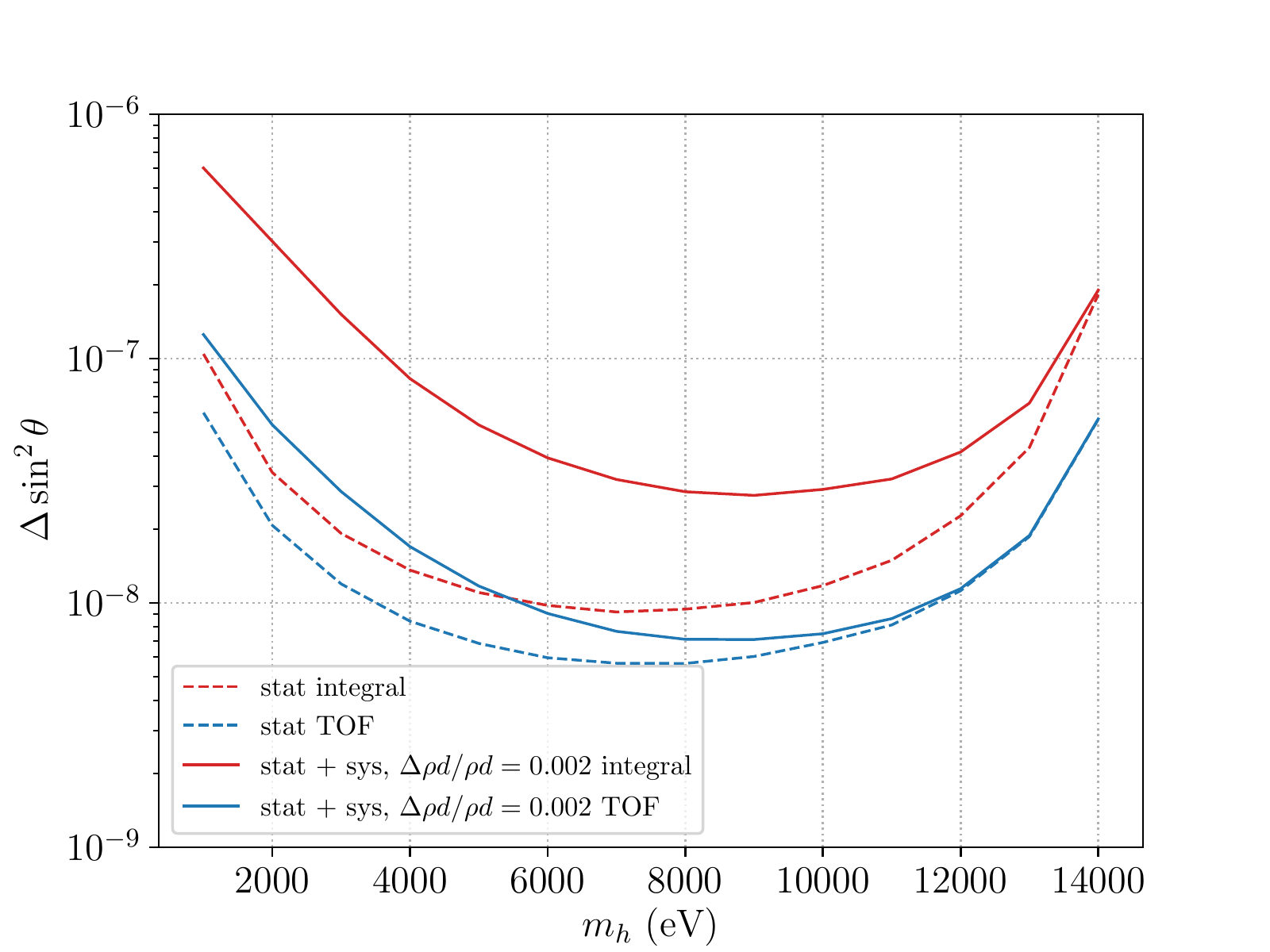}
  \caption{Sensitivity (1 $\sigma$) of ideal TOF mode (blue) compared with integral mode (red). Both statistical uncertainty (dashed lines) and combined uncertainty with exemplary systematics (full lines) in form of column density uncertainty $\Delta\rho d / \rho d = 0.002$ affecting the inelastic scattering cross section in the WGTS. It can clearly be seen that the sensitivity gain by a TOF mode is especially significant if the uncertainty of the column density is accounted for. The results are based on three years measurement time, distributed uniformly on the retarding potential within an interval of [4; 18.5] \si{keV} with steps of \SI{0.5}{keV}.}
  \label{fig:idealtofsensitivity}
\end{figure}

An exemplary fit is shown in \Figref{fig:sterilefit} for a sterile neutrino with mass $m_h = \SI{2} {keV}$ and a mixing of $\sin^2\theta=10^{-6}$ assuming an ideal TOF measurement and using four exemplary retarding potentials of 15, 16, 17 and 18 \si{keV}. In this case the sterile neutrino mass has not been fixed but used as a free fit parameter to test the ability to fit the sterile neutrino mass, given a sufficiently high active-sterile mixing. While it is in principle sufficient to use only one retarding potential closely below the sterile neutrino kink, in practice a multitude of retarding potentials is necessary. The reasons are that, on one hand, the mass of the sterile neutrino is unknown and, on the other hand, that it is also necessary in order to determine the other parameters. In contrast to the pure sterile active mixing sensitivity estimation (\Figref{fig:idealtofsensitivity}) the heavy neutrino mass has been used as free fit parameter. It shows that the method is capable of a sensitive mass determination as well, in case the mixing angle is large enough. However, since most parts of the sensitive regions of the TOF method are disfavored by X ray satellite measurements \cite{Watson2012}, it seems unlikely that a mass fit will be possible. 

\begin{figure}
  \includegraphics[width=1.05\linewidth]{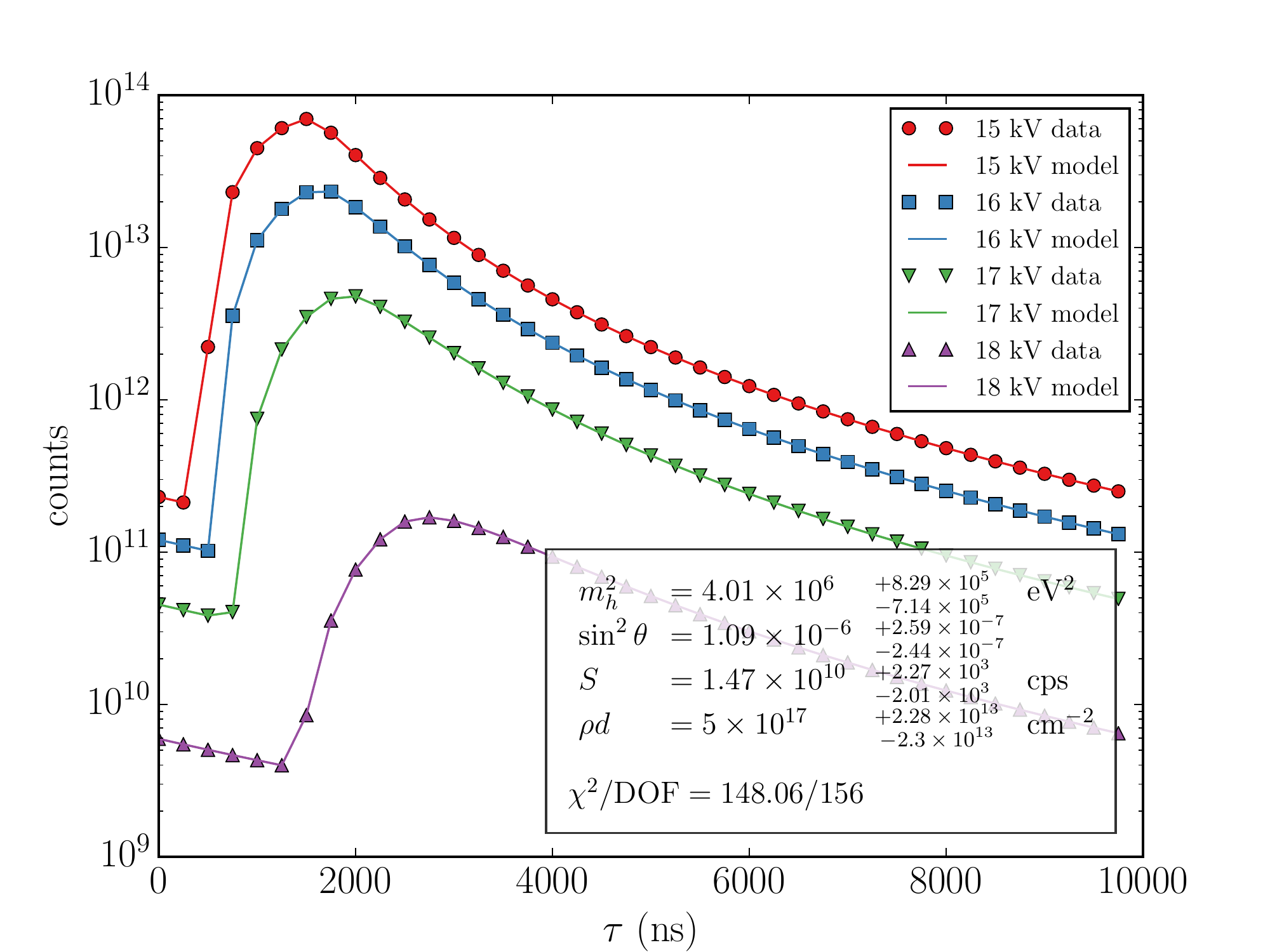}
  \caption{Exemplary fit of a sterile neutrino with mass $m_h = \SI{2} {keV}$ and low mixing $\sin^2\theta=10^{-6}$ (not visible by eye), assuming an ideal TOF measurement and using four exemplary retarding potentials of 15, 16, 17 and 18 \si{keV}. The fit includes the systematic uncertainty of the column density $\rho d$, as well as the sterile neutrino mass as free fit parameters. The overall count rate increases with decreasing retarding potential. }
  \label{fig:sterilefit}
\end{figure}

\subsection{Optimization and Integrity}

\subsubsection*{SCAMC Variance}

\begin{figure}
  \includegraphics[width=1.1\linewidth]{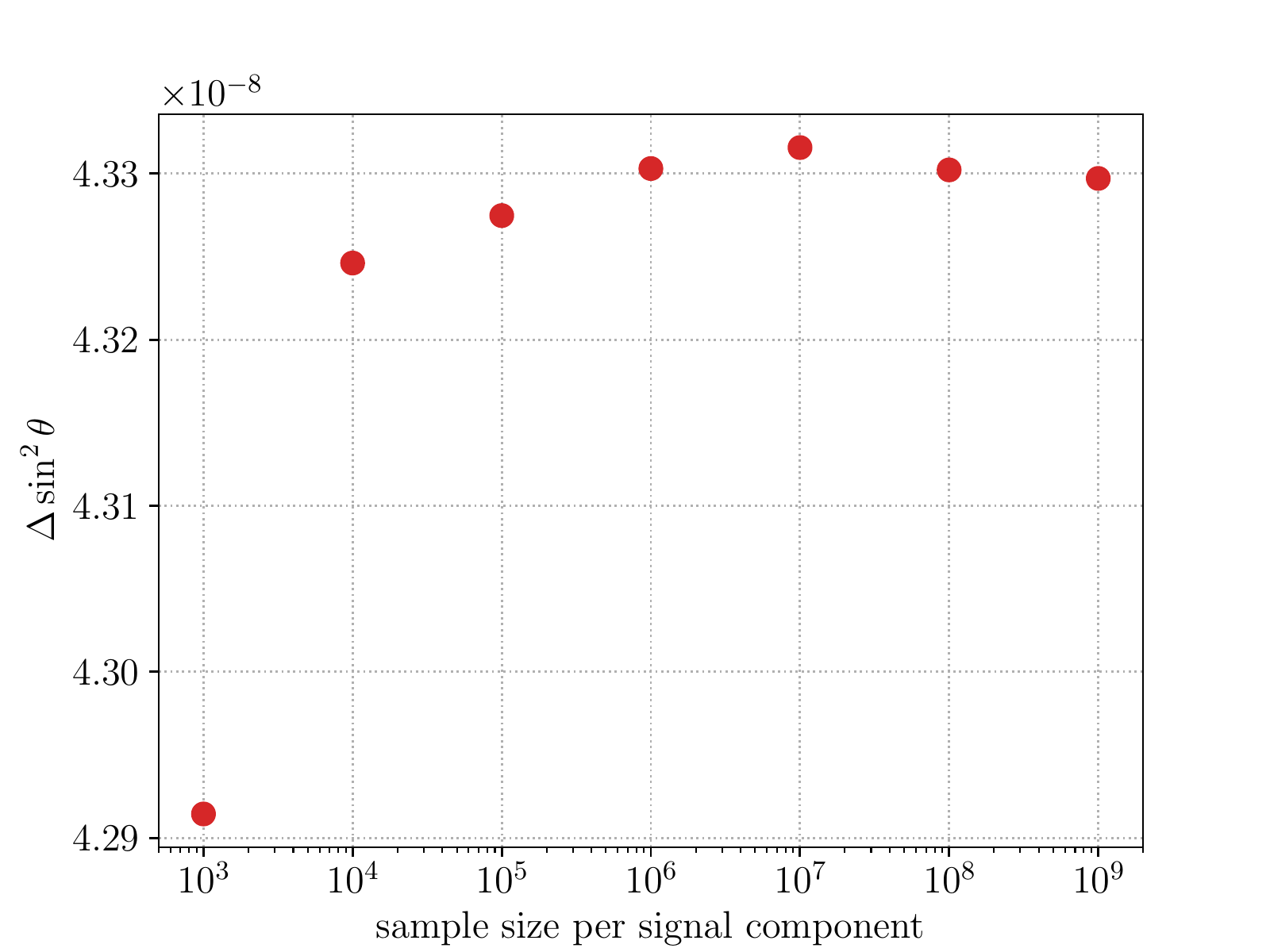}
  \caption{Estimated statistical sensitivity with ideal TOF mode for a \SI{2}{keV} neutrino as a function of the MC sample size per signal component (see \ref{app:decomposition}, \Eqref{eq:subsignals}), using a measurement interval of [4; 18.5] \si{keV} with steps of 0.5 keV. The total signal sample size per TOF spectrum is given by \Eqref{eq:totalsignalsize}. The background has been simulated using 10 times the signal component sample size.}
  \label{fig:sizecomp}
\end{figure}

In order to show that the SCAMC method is really working as expected, it has been tested using different Monte Carlo sample sizes. A necessary condition is convergence of the result towards a constant value with growing sample size. As described in \ref{app:decomposition}, the signal itself is split into signal components defined by slices of the signal TOF spectrum in $m_h$-space (\Eqref{eq:subsignals}). \Figref{fig:sizecomp} shows the ideal TOF mode statistical sensitivity for a \SI{2}{keV} neutrino as a function of the sample size used for each such component of the signal. The components have been sampled with steps of $\SI{0.1}{keV}$ in terms of $m_h$. The total signal sample size per TOF spectrum is thus given by 

\begin{equation}
  N_\up S = N_\up C \cdot \frac{(E_0 - m_h - qU)}{\SI{0.1}{keV}} \ ,
  \label{eq:totalsignalsize}
\end{equation}

\noindent  where $N_\up C$ denotes the sample size per component. For minimum $qU$ and $m_h$ it amounts to $\sim 150 \cdot N_\up C$. The background has been simulated with a sample size of $10 \cdot N_\up C$.

It can be seen that convergence is met and that with sub-sample sizes such as $10^4$ per component the expected result is approximated with less than 1 percent uncertainty.

\subsubsection*{Measurement Interval}

\begin{figure}
  \includegraphics[width=1.05\linewidth]{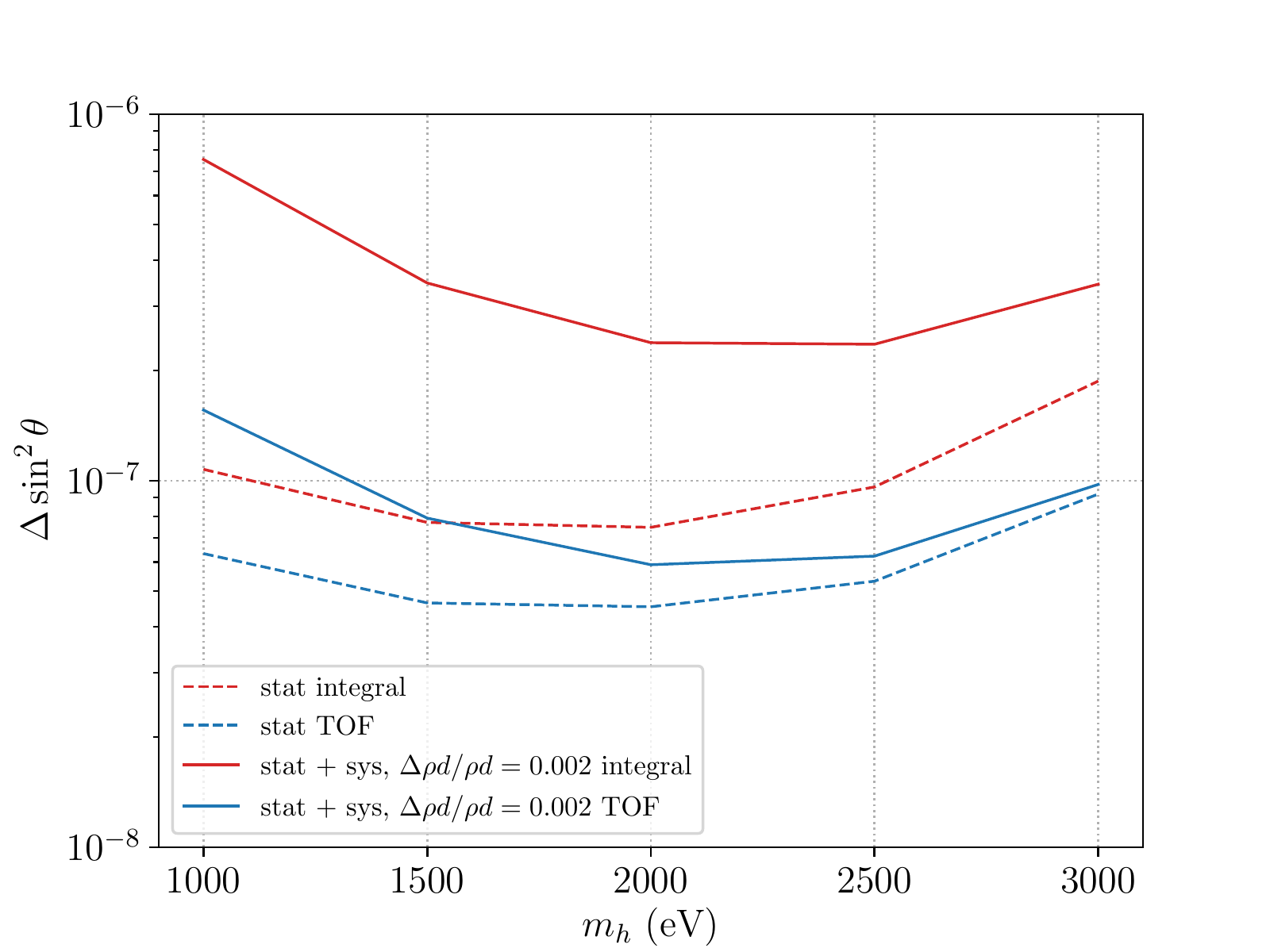}
  \caption{Same settings as in \Figref{fig:idealtofsensitivity} but with a measurement interval of [15; 18.5] \si{keV}. The narrowing of the measurement interval shows no benefit even if the sterile neutrino kink is within the interval.}
  \label{fig:idealtofsensitivity2kev}
\end{figure}

\Figref{fig:idealtofsensitivity2kev} shows the sensitivity to $\sin^2\theta$ in a similar way as \Figref{fig:idealtofsensitivity}, but for a measurement interval of [15; 18.5] \si{keV}, roughly centered around a 2 keV neutrino, as favored in Ref. \cite{Destri2013b}. It can be seen in comparison that there is no benefit of restricting the measurement interval to a narrow region in search for a sterile neutrino with a given energy. This seems counter-intuitive at first, but is has to be kept in mind that the sterile neutrino signal is not localized at the kink, but instead contributes to the whole spectrum below. In contrast to dedicated 'kink-search' methods \cite{Mertens2015a}, all spectral parts contribute to the sensitivity in a $\chi^2$ fit. While the relative difference made by a sterile neutrino signal might be smaller at lower retarding potentials, this drawback is however balanced by a larger count-rate at lower potentials.

\subsubsection*{Measurement Step Size}

\begin{figure}
  \includegraphics[width=1.1\linewidth]{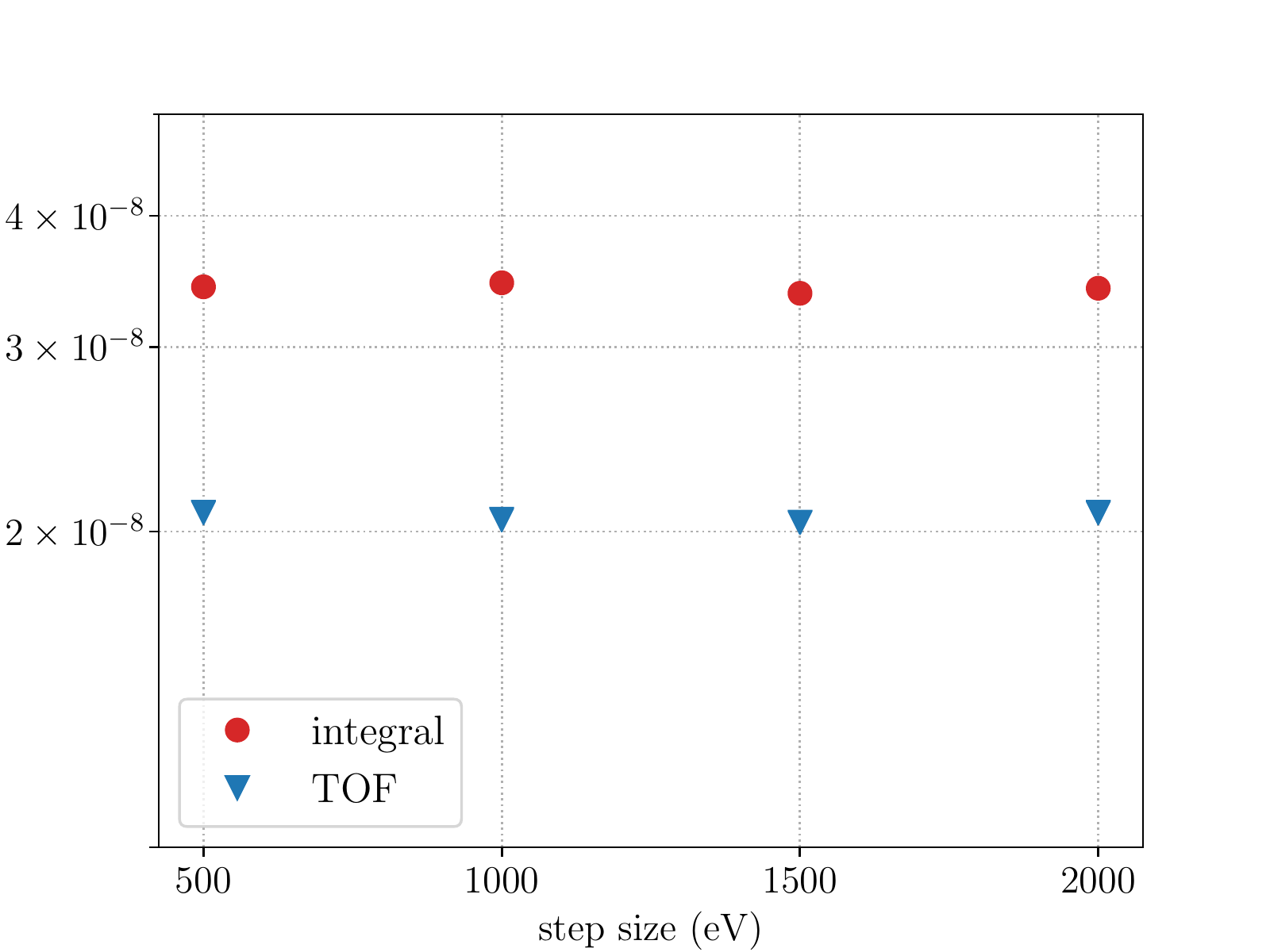}
  \caption{Statistical sensitivity as a function of the step size between measurement points of the retarding potential $qU$ for a sterile neutrino with $m_h = \SI{2}{keV}$, with constant total sample size. The measurement interval is [4; 18.5] \si{keV} for a total measurement time of three years. }
  \label{fig:stepcomp}
\end{figure}

\Figref{fig:stepcomp} shows the statistical sensitivity as a function of the spacing between different measurement points of the retarding potential $qU$. The total sample size has been kept constant. The simulations show no preference towards any particular value. That appears unintuitive, since one would expect a narrower spacing to have beneficial effects on a distinct kink search. Yet, as mentioned in the last paragraph, the sterile neutrino signal is not localized, but manifests itself in relative count rate differences between the measurement points with a spectral feature as broad as the mass of the sterile neutrino $m_h$. Therefore, a larger step size does not weaken the sensitivity in principle because the measurement time is distributed over less points. Anyway, it is in general recommended to use a step size lower then the smallest possible heavy neutrino mass, since otherwise it is possible that there are not enough vital measurement points above the kink.

The benefit of a TOF measurement can be explained in this context as follows: TOF spectra carry extra information about the differential energy distribution closely above each measurement point. That equates to knowledge about the slope of the integral spectrum at these measurement points.

\subsection{Gated Filter Sensitivity}

\Figref{fig:gatedspectrum} shows exemplary TOF spectra using Gated Filtering (GF, see \Figref{fig:gatedfilter}). It illustrates how GF works: without the gate (cyan points), the arrival time spectrum is isochronous. However, with activated gate, a certain portion is cut away from the isochronous spectrum. For a given repetition time $t_r$ and duty cycle $\xi$, the duration in which the gate is open is given by $t_r \cdot \xi$. The GF arrival time filter thus is smeared with a step function when compared to the raw TOF spectrum. Reducing the duty cycle $\xi$ makes the arrival time spectrum approximate the TOF spectrum of \Figref{fig:sterilefit}, however with a loss of overall rate. Electrons with a TOF greater than the repetition time $t_r$ lead to the wrongful attribution of the corresponding events to a later period, which can be seen in the first few bins. However, since TOF spectra at several keV below the endpoint are rather sharp, this effect is small for repetition times of $\sim \SI{10}{\micro s}$.

\begin{figure}
  \includegraphics[width=1.05\linewidth]{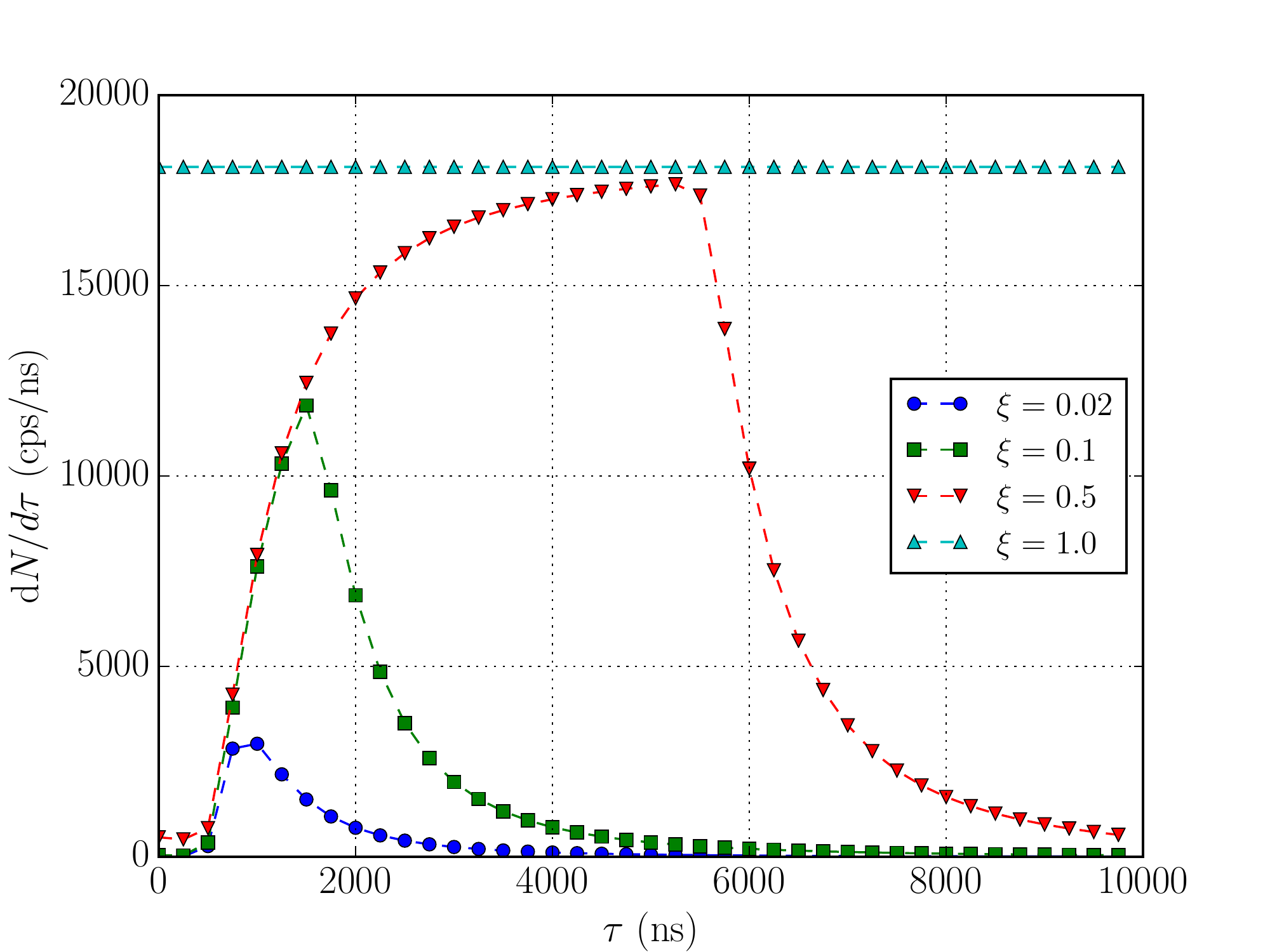}
  \caption{Exemplary Gated Filter arrival time spectra for different duty cycles. Retarding energy is $qU=\SI{17}{keV}$ and repetition time $t_r = \SI{10}{\micro s}$. The active-sterile mixing has been set to $\sin^2 \theta=0$. Activating the gate and decreasing the duty cycle cuts away portions of the arrival time spectrum, which is isochronous without gate.}
  \label{fig:gatedspectrum}
\end{figure}

\Figref{fig:gatedsensitivity} shows the sensitivity for two exemplary gated filter scenarios with a constant duty cycle of 0.1. The scenario is based on the assumption that the existing focal plane detector (FPD) of KATRIN is used, which is optimized for a measurement near the endpoint of the $\upbeta$ spectrum and thus can not maintain much higher count-rates. The bottleneck is particularly the per-pixel rate which should not exceed $\sim\SI{e3}{cps}$ within a window of some $\si{\micro s}$. This limitation holds for the current data acquisition and might be improved in the future. In this simulation, an exemplary overall reduction of the signal rate by a factor $10^5$ has been chosen which will ensure a flux compatible with the current hardware. Since the gated filter periodically blocks the flux of electrons, the rate can be increased again with respect to the integral mode. The actual allowed rate with the gated filter depends on the readout electronics and will effectively be between two extremal values. In an optimistic case, short-time excess of the rate is tolerable, while the average rate has to be at the same level as with the integral mode. In a conservative case, also short-time excess leads to pile-up, which means that instead the peak rate may not exceed the constant rate of the integral mode (see \Figref{fig:gatedspectrum}). The repetition rate has been fixed at \SI{10}{\micro s}, which will ensure coverage the vast part of the TOF spectrum. The measurement interval has been limited to [15; 18.5] \si{keV} since it is not believed to be viable to pulse the pre-spectrometer more than several keV.

\begin{figure}
  \includegraphics[width=1.05\linewidth]{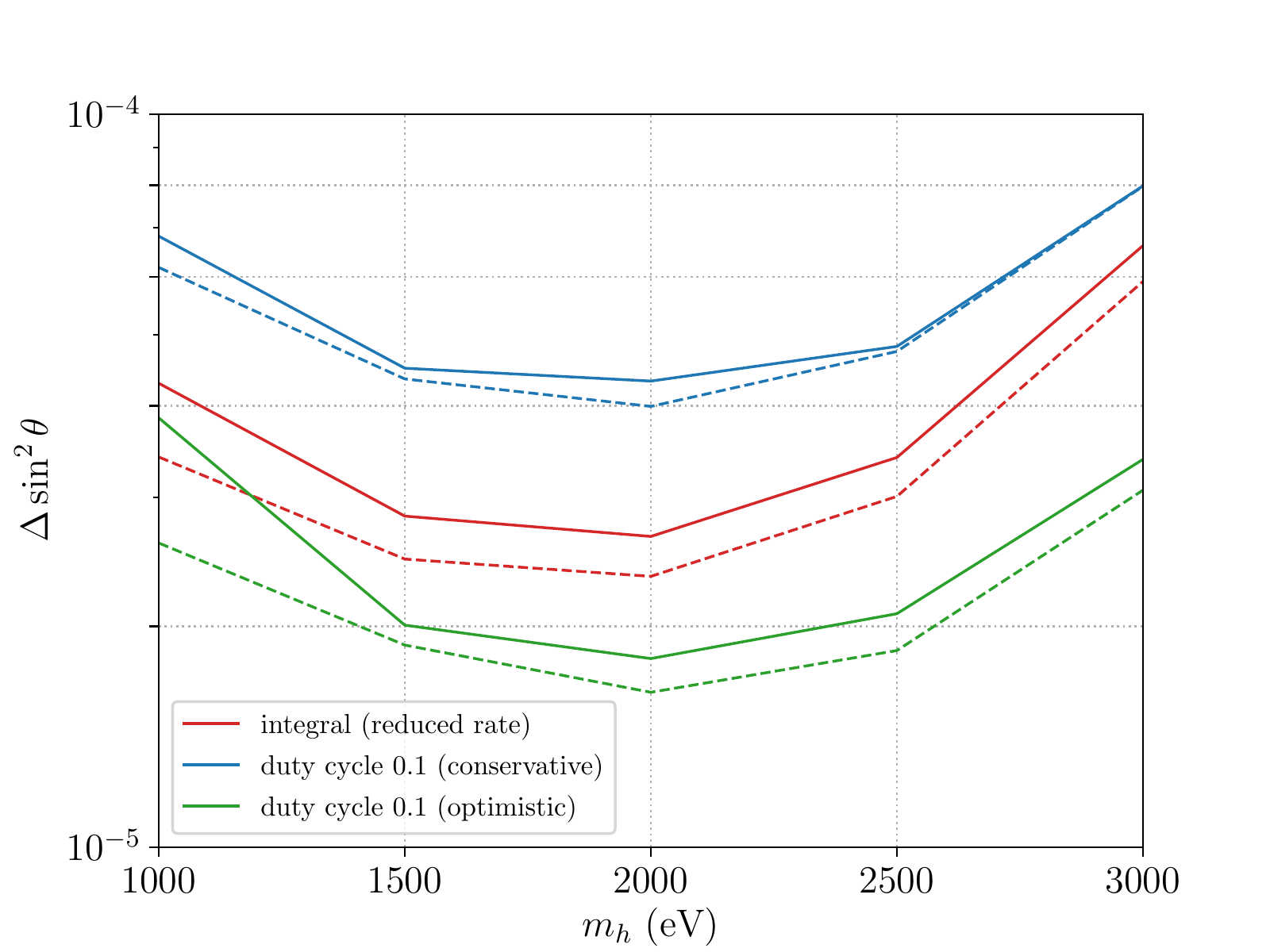}
  \caption{Sensitivity (1 $\sigma$) of integral mode with the rate reduced by a factor \num{1e5} (red), compared with a conservative gated filter TOF scenario (blue) with same peak rate as the integral mode and an optimistic gated filter TOF scenario (green) with the same total rate as the integral mode. Both statistical uncertainty (dashed lines) and combined statistical plus systematic uncertainty including a column density uncertainty $\Delta\rho d / \rho d = 0.002$ (full lines, see section \ref{sec:systematics}) are plotted. The duty cycle is 0.1 for both gated filter scenarios. Measurement interval has been [15; 18.5] \si{keV} for three years data taking.  The repetition time is $t_r = \SI{10}{\micro s}$ for all retarding potentials.   }
  \label{fig:gatedsensitivity}
\end{figure}

It can be seen that in the the gated filter beats the integral mode in the optimistic case, but not in the conservative case. This means that the loss of rate in the conservative case is too high to be compensated by the additional TOF information. In case the detector readout electronics sufficiently tolerates short-time excesses, the loss of statistics by the gated filter can, nevertheless, be compensated and additional TOF information is gained. Note, however, that in the scenario of an upgraded future detector which tolerates the full rate from the tritium source, the integral mode outperforms the gated filter mode, since there is now way in this case to increase the rate further.



\section{Summary and Discussion}

It has been shown that TOF spectroscopy in a KATRIN context is in principle able to boost the sensitivity of the sterile neutrino search significantly. \Figref{fig:idealtofsensitivity2kev} suggests an improvement of up to a factor two in terms of pure statistical uncertainty down to at maximum $\sin^2\theta \sim \num{5e-9}$ for a sterile neutrino of $m_h=\SI{7}{keV}$ at one $\sigma$. If the exemplary systematic uncertainty of the inelastic scattering cross section is considered, the sensitivity is only mildly weakened in contrast to the integral mode, which is in that case outperformed by the TOF mode by up to a factor five. However, the practical realization of a sensitive TOF measuring method is still work in progress. Given the current hardware, which requires a reduction of the signal rate, the gated filter method might be able to realize a TOF mode with a slight sensitivity increase compared to the integral mode under the condition that the detector tolerates short-time excesses of the rate and that it is possible to ramp the pre-spectrometer potential some keV within $\sim \SI{0.1}{\micro s}$. From a long-term point of view, the concept of an upgraded differential detector \cite{Mertens2015} which is capable of extreme rates up to \SI{e10}{cps} is very promising. If there is sufficient progress in developing a sensitive TOF measurement method, a beneficial strategy could be a combined measurement to eliminate systematics and perform cross-checks.

\begin{acknowledgements}

We would like to thank S. Enomoto for discussions. This work is partly funded by BMBF under contract no. 05A11PM2 and DFG GRK 2149.  

\end{acknowledgements}

\appendix

\section{Unchanged \texorpdfstring{$\chi^2$}{Chi Square} Properties with SCAMC}
\label{app:proof}
In the following it is shown that the properties of the $\chi^2$ function defining the sensitivity, which are position and width of the minimum with respect to any parameter of interest, are independent of the choice of the background model $\Phi_\up B'$. This works as well for a Poissonian log-likelihood, but for brevity we show it on a $\chi^2$ example. First we define the expectation value for the $i$-th bin,

\begin{equation}
  \lambda'_i = \lambda_{\up S_i} + \lambda'_{B_i} = n \ (c_\up S \Phi_{\up S_i} + c_\up B \Phi'_{B_i}) \ ,
\end{equation}

\noindent using the definition of the approximated model \eqref{eq:approxmodel}, and assume that the background prediction $\lambda'_{B_i}$ is independent of the parameter of interest $\mu$,

\begin{equation}
  \frac{\mathrm d}{\mathrm d \mu} \lambda_{B_i}' = 0 \ .
\end{equation}

For the proof we differentiate $\chi^2$ with respect to $\mu$ and demand that the result is approximately independent of the choice of the background model $\Phi_\up B'$:

\begin{align}
	\chi^2 (\mu) = & \sum_i \frac{(n_i - \lambda_i'(\mu))^2}{\lambda_i'(\mu)} \\
	\frac{\mathrm d}{\mathrm d \mu} \chi^2 = & \sum_i \frac{\lambda_i'\frac{\mathrm d}{\mathrm d \mu} (n_i - \lambda_i')^2 - (n_i - \lambda_i')^2\frac{\mathrm d}{\mathrm d \mu} \lambda_i'}{\lambda_i'^2} \nonumber\\
	 = & \sum_i \frac{-2\lambda_i'(n-\lambda_i')\frac{\mathrm d}{\mathrm d \mu} \lambda_{\up S_i} - (n-\lambda_i')^2\frac{\mathrm d}{\mathrm d \mu}  \lambda_{\up S_i}}{\lambda_i'^2}\nonumber\\
	 = & - \sum_i \frac{(n_i^2-\lambda_i'^2) \frac{\mathrm d}{\mathrm d \mu}  \lambda_{\up S_i} }{\lambda_i'^2} \nonumber\\
	 = & \sum_i \left(1 - \frac{n_i^2}{\lambda_i'^2}\right)\frac{\mathrm d}{\mathrm d \mu}  \lambda_{\up S_i} \nonumber\\
		 = & \sum_i \left(1 - \frac{n_i^2}{(\lambda_{B_i}' +\lambda_{\up S_i} )^2}\right)\frac{\mathrm d}{\mathrm d \mu}  \lambda_{\up S_i}\nonumber \\
		 = & \sum_i \left(1 - \left(\frac{\lambda_{B_i}'}{n_i} + \frac{\lambda_{\up S_i}}{n_i} \right)^{-2} \right)\frac{\mathrm d}{\mathrm d \mu}  \lambda_{\up S_i} 
\end{align}

\noindent The variable $n_i$ is Poisson distributed with mean $\lambda'_i(\mu_0)=\lambda_{\up S_i}(\mu_0)+\lambda'_{B_i}$, where $\mu_0$ is the null-hypothesis for $\mu$. Due to self-consistency, $\frac{\lambda_{B_i}'}{n_i}$ is approximately independent from the choice of $\Phi_\up B'$, as long as the order of magnitude is in agreement $\Phi'_B \sim \Phi_\up B$. The latter condition ensures that the Poissonian uncertainty of $n_i$, which is given by $\sqrt{\lambda'_i(\mu_0)}$, is approximately correct. \qed

Note that the proof is only correct in the simplified case of one parameter of interest and no correlation with nuisance parameters. However, the simulation results in this paper show that there is valid reason to expect the method to work also for more complex problems as long as there is no heavy parameter correlation.

\section{Sterile Neutrino Mass Decomposition of TOF Spectra}
\label{app:decomposition}

The simulation of the TOF spectra has further been optimized with the aim of being able to use the sterile neutrino mass $m_l$ as a free parameter with a minimum of computational overhead. The idea is to decompose the sterile neutrino components of the TOF spectra, $\Phi_\up S$, into sub-spectra $\Phi_{S_k}$ which can be added subsequently to obtain the signal for a given sterile neutrino mass $m_h$. That works as follows: at first a number $J$ of grid points with heavy neutrino masses $m_j$ are chosen. For each grid-point $j$, the signal spectrum is given as the sum of all sub-signals from $j$ up to $J$, 

\begin{align}
	\Phi_\up S (m_j) = \sum_{k=j}^{J}  \Phi_{S_k} \ .
\end{align}

\noindent The sub-signals $\Phi_{S_k}$ constitute the difference of two TOF spectra with adjacent sterile neutrino masses. The total TOF spectrum for the sterile component can then be written as 

\begin{align}
\frac{\mathrm d N}{\mathrm d \tau} (m_j) & =  \frac{\mathrm d N}{\mathrm d \tau} (m_J) +  \sum_{k=j}^{J-1} \bigg( \frac{\mathrm d N}{\mathrm d \tau} (m_k) - \frac{\mathrm d N}{ \mathrm d \tau} (m_{k+1}) \bigg) \ .
	\label{eq:subsignals}
\end{align}

Each sub-component in the sum will be sampled separately. The difference between two TOF spectra can be sampled just like any TOF spectrum, as outlined, by replacing the $\upbeta$-spectrum in \eqref{eq:tofspec} also with the difference of two $\upbeta$ spectra corresponding to the neutrino masses $m_k$ and $m_{k+1}$. Via \eqref{eq:subsignals}, that gives then the sterile contribution of the TOF spectrum for each mass value $m_j$ on the grid. For sterile neutrino masses between the grid points, the resulting spectrum is then calculated by cubic spline interpolation. The strategy is illustrated in \Figref{fig:subsignals}. 

\begin{figure}
  \includegraphics[width=\linewidth]{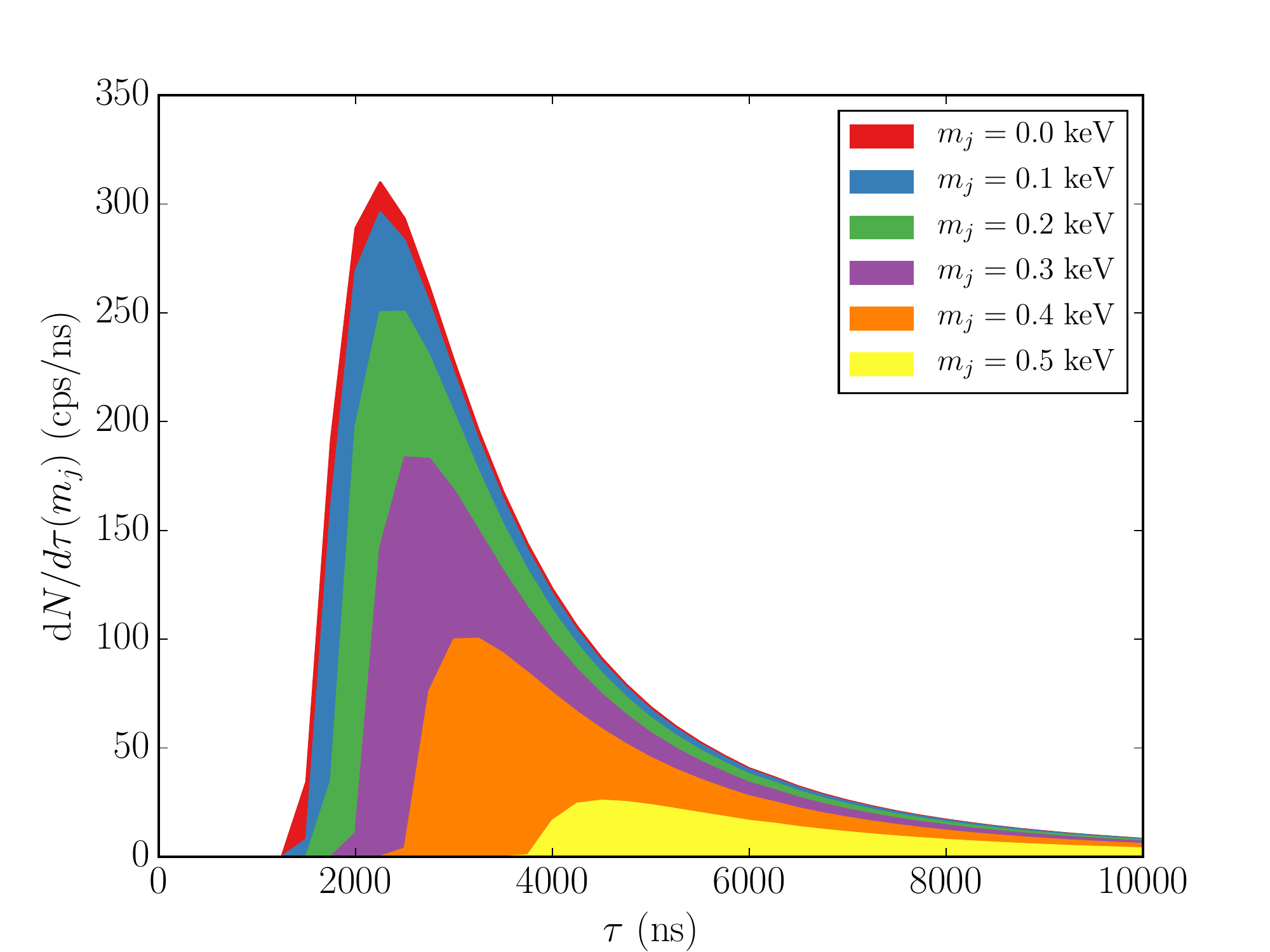}
  \caption{Illustration of the calculation of sterile component of the electron TOF spectrum via subsequent addition of sub-components according to \eqref{eq:subsignals}. The figure shows the sterile components of the TOF spectrum \eqref{eq:steriletof} for different sterile neutrino masses $m_j$ on a grid for a retarding potential of $qU=\SI{18}{kV}$. Each colored area corresponds to a sub-component between two adjacent mass values. The component for any sterile neutrino mass $m_j$ is then given by the sum of all areas below the envelope.}
  \label{fig:subsignals}
  \end{figure}

In addition to the reuse of already simulated Monte Carlo events, this strategy has the possible advantage of a smoother interpolation in bins with small statistics, which are possible for high flight times $\gtrsim \SI{40}{\micro s}$. By the de-composition and subsequent addition of the components, monotony between the interpolation grid points is guaranteed. However, if a sufficient overall sample size is chosen, this effect should not matter significantly. 

\bibliographystyle{spphys}
\bibliography{wdmtof.bib}

\end{document}